\documentclass[sigconf]{acmart}

\renewcommand\footnotetextcopyrightpermission[1]{}

\usepackage{multicol}
\usepackage{stfloats}
\usepackage{multirow}
\usepackage{booktabs}
\usepackage{tcolorbox}
\usepackage{soul}
\usepackage{balance}
\usepackage{tcolorbox}
\usepackage{makecell}
\usepackage{diagbox}
\usepackage{graphicx}

\AtBeginDocument{%
  }

\acmISBN{978-1-4503-XXXX-X/2018/06}

\begin{document}

\title[What Characterizes a Software Leader?]{What Characterizes a Software Leader? \\ Identifying Leadership Practices from Practitioners' Social Media}

 \author{Murilo Coelho}
 \authornote{Corresponding author}
 \affiliation{%
   \institution{Atlantico Institute}
   \institution{State University of Ceará}
   \city{Fortaleza}
   \state{Ceará}
   \country{Brazil}
 }
 \email{paulo\_coelho@atlantico.com.br} 

 \author{Denivan Campos}
 \affiliation{%
   \institution{University of Molise}
   \city{Pesche}
   \country{Italy}
 }
 \email{denivan.dasilva@unimol.it}

 \author{Mariana Maia Bezerra}
 \affiliation{%
   \institution{Atlantico Institute}
   \city{Fortaleza}
   \state{Ceará}
   \country{Brazil}
 }
 \email{mariana\_maia@atlantico.com.br}

 \author{Matheus Paixao}
 \affiliation{%
   \institution{State University of Ceará}
   \city{Fortaleza}
   \state{Ceará}
   \country{Brazil}
 }
 \email{matheus.paixao@uece.br}

 \author{Allysson Allex Araújo}
 \affiliation{%
   \institution{Federal University of Cariri}
   \city{Juazeiro do Norte}
   \state{Ceará}
   \country{Brazil}
 }
 \email{allysson.araujo@ufca.edu.br}

 \author{Sávio Freire}
 \affiliation{%
   \institution{Federal Institute of Ceará}
   \institution{State University of Ceará}
   \city{Morada Nova}
   \state{Ceará}
   \country{Brazil}}
 \email{savio.freire@ifce.edu.br}

\renewcommand{\shortauthors}{Coelho et al.}

\begin{abstract}

\textbf{Context:} Leadership has been extensively studied in management and agile software development; however, prior research predominantly focuses on formal roles and predefined leadership models, offering limited insight into how leadership is experienced and demonstrated by software practitioners in everyday practice.
\textbf{Objective:} Our goal is to identify and categorize leadership practices as perceived and reported by software development practitioners based on their professional experiences. 
\textbf{Method:} We conducted a content analysis of 116 practitioner-authored articles published on the Dev.to online community. Articles were systematically collected, screened, and coded, resulting in the extraction, correlation analysis and  categorization of leadership practices grounded in practitioners’ narratives.
\textbf{Results:} We identified 103 practices for software project leaders, distinguished between recommended and discouraged ones. These practices were organized into five categories: People Management \& Development, Processes \& Execution, Professional \& Personal Growth, Communication \& Articulation, and Strategic Vision. The most recurrent recommended practices include \textit{Cultivating \& Practicing Interpersonal Skills}, \textit{Managing \& Delegating Team Work}, and \textit{Practicing \& Developing Managerial Skills}, whereas \textit{ Micromanagement}, \textit{Counterproductive Work Patterns}, and \textit{Counterproductive Communication Styles} emerged as the most frequent discouraged practices. We organized all practices into a conceptual map.
\textbf{Conclusion:} The findings indicate that software leadership is mainly associated with managerial and interpersonal practices rather than technical expertise. The resulting conceptual map summarizes these practices and can serve as a reference for understanding leadership in software development contexts.
\end{abstract}

\ccsdesc[500]{General and reference~Empirical studies}
\ccsdesc[500]{Social and professional topics~Software management}

\keywords{Leadership, Social Media, Human Aspects, Software Engineering.}

\maketitle
\section{INTRODUCTION}

In the dynamic context of software engineering (SE), project success extends beyond technical proficiency \cite{procaccino2005software}. 
Soft skills, particularly leadership, are fundamental in shaping collaboration, maintaining motivation, and ensuring the sustainability of software teams \cite{li2015makes, capretz2014bringing, coelho2024estudo, kalliamvakou2017makes, coelho2025soft}. 
Leadership extends beyond a formal role, functioning as a process of social influence in which an individual guides and facilitates a team toward shared goals. \cite{house1975path,li2012leadership,subramanian2024towards,salas2005there, xu2024impact}. 
Effective leadership enables alignment, autonomy, and trust, turning individual expertise into collective achievement \cite{gren2022makes}. 
On the other hand, inadequate leadership may undermine psychological safety and compromise project outcomes \cite{piwowar2025sustainability}.

In SE, leadership manifests itself through different dimensions, one of the most relevant being technical leadership, whose role is to guide decisions related to system design and architecture, ensuring code quality, technical consistency, and the alignment of engineering practices with organizational objectives \cite{faraj2006leadership, adanigboconceptual2025}. Comprehending leadership in this domain is an essential part of improving technical performance and team dynamics in software development. 

Although leadership is widely recognized as a critical factor in software projects \cite{thite1999leadership, khattak2025unwrapping}, most prior work has examined leadership through the lens of managerial or agile frameworks, emphasizing formal roles such as Scrum Master or Project Manager \cite{xu2018role, gren2022makes}. 
These studies tend to focus on methodological coordination and role responsibilities, often overlooking the informal and practice-based aspects of leadership. 
Consequently, there remains limited empirical understanding of what leadership looks like in everyday software work and how practitioners themselves describe or evaluate leadership behaviors \cite{modi2020leadership}. 
This gap constrains our ability to connect theoretical constructs of leadership with the realities of software practice, despite evidence that effective leadership strongly influences team performance and overall project success \cite{gren2017group}.

To address this gap, we aim to investigate leadership as it is constructed and reflected by software practitioners in social media platforms. 
They encompass blog posts and reflective essays that capture how professionals articulate and share their experiences \cite{schopfel2010grey,KitchenhamGrey}. 
Analyzing this content allows researchers to access unsolicited practitioner perspectives, minimizing researcher bias while capturing community-level sensemaking about leadership in real-world settings \cite{Garousi2020GreyLiterature}. 

In this study, we selected the \hyperlink{https://dev.to}{Dev Community} (Dev.to) platform as the empirical source. 
Dev.to is an open-access global community focused on dialogue among software developers, making it a suitable environment for studying leadership as a socially constructed phenomenon \cite{papoutsoglou2021mining}. Previous studies have successfully approached Dev.to to explore different SE topics \cite{oliveira2025investigating, Cerqueira2025TOSEM}. 
Using Dev.to’s official API, we retrieved 1,815 articles tagged with \textsc{leadership} and other related keywords. After automated filtering and a two-phase manual screening, 116 curated articles were retained for analysis.

We applied content analysis \cite{Mayring_2000} to extract and organize leadership practices mentioned in the analyzed articles. The process yielded 103 distinct practices, classified as either recommended or discouraged behaviors. Through open coding \cite{strauss1998basics}, these practices were further analyzed along two complementary perspectives. The first captured the \textbf{nature of the practices}, distinguishing between technical actions related to software development activities and managerial behaviors associated with coordination and team support. The second perspective grouped practices into five categories in relation to \textbf{recommended and discouraged practices}: People Management \& Development, Strategic Vision, Processes \& Execution, Communication \& Articulation, and Professional \& Personal Growth. 
We further analyzed the practices using a quantitative approach. We calculated the number of mentions for each practice to assess their prominence in the discourse and applied statistical analyses to examine co-occurrence patterns. 
All datasets, coding schemes, and analytical scripts are publicly available in an open replication package \cite{repo}.

This study makes three major contributions to the understanding of software leadership.
First, it offers an empirically grounded characterization of leadership practices derived from practitioners’ discourse in social media, capturing experience-based reflections shared on Dev.to that extend beyond formal leadership roles and organizational contexts.
Second, it introduces a structured conceptual map that organizes 104 identified practices (both recommended and discouraged) into five categories.
Third, it translates these findings into actionable findings for researchers and practitioners. For researchers, it provides a community-informed empirical basis for theorizing about leadership behaviors in software projects. For practitioners, it highlights concrete practices that can guide leadership development, team mentoring, and organizational training initiatives grounded in real-world experience.
\section{RESEARCH METHOD} \label{sec:researchmethod}
The objective of this study, following the Goal–Question–Metric (GQM) paradigm \cite{caldiera1994goal,wohlin2012experimentation}, is defined as follows: To \textit{analyze} leadership practices in software development projects, \textit{with the purpose of} identifying, characterizing and categorizing both recommended and discouraged practices, \textit{from the perspective} of software practitioners, \textit{in the context} of the Dev.to. To achieve this objective, we formulated the following research questions (RQs):

\begin{itemize}
    \item \textbf{RQ1:} \textit{What are the recommended practices for leaders in software development projects?} \textit{Rationale}: Our aim is to identify the actions and behaviors that are considered effective or advisable for leaders in the context of software teams.
    \item \textbf{RQ2:} \textit{What are the discouraged practices for leaders in software development projects?} \textit{Rationale}: We seek to identify the actions and behaviors that are considered detrimental or to be avoided for leaders in the context of software teams.
\end{itemize}

The following subsections presents the data collection and analysis procedures we employed to answer the RQs. These procedures were based on previous studies that investigated distinct SE phenomena in social media platforms, such as Stack Exchange~\cite{Gama2020,santanaSBES,santanaIEEESW,freireREFSQ,gomes2023investigating,felipeUFBA} and Dev.to~\cite{oliveira2025investigating,cerqueira2023thematic,cerqueira2024IEEESW,Cerqueira2025TOSEM}.

\vspace{-0.1cm}

\subsection{Data Collection}

For this study, we used social media as the data source. This type of data provides access to practitioner perspectives and helps bridge the gap between academic research and the day-to-day practices of software professionals \cite{KitchenhamGrey,schopfel2010grey}. In particular, social media content reflects diverse experiences and viewpoints, offering a complementary lens to traditional empirical methods for identifying leadership practices grounded in real-world contexts.
As previously mentioned, we selected the Dev.to for data collection, as it is a global community where software professionals publish articles, share reflections, and exchange feedback on technical, organizational, and social aspects of software work. Its open and interactive environment fosters peer learning and discussion on human and collaborative factors in software development. The platform has been widely used in empirical SE studies addressing human and social dimensions of professional practice \cite{papoutsoglou2021mining, cerqueira2023thematic, cerqueira2024IEEESW, Cerqueira2025TOSEM, oliveira2025investigating}.

Data collection was conducted in August 2024, which served as a cut-off date for this study. 
We used \hyperlink{https://developers.forem.com/api}{Forem}, the official API for the Dev.to, to programmatically search for articles based on tags.
The process was conducted in two stages. First, we performed a search using the \textsc{\#leadership} tag, which resulted in 1,659 articles.
To maximize the retrieval of relevant articles, we then expanded the search. 
The strategy consisted of selecting a set of associated tags that contained variations of the root word \textit{lead}. 
Thus, the following 22 tags were included: \textsc{\#effectiveleadership}, \textsc{\#leadershipskills}, \textsc{\#leadershipdevelopment}, \textsc{\#careerladder}, \textsc{\#engineeringleaders}, \textsc{\#engineeringleadership}, \textsc{\#leadbyexample}, \textsc{\#leaddev}, \textsc{\#leaddeveloper}, \textsc{\#leadengineer}, \textsc{\#leader}, \textsc{\#leaders}, \textsc{\#leadgenratio}, \textsc{\#productleader}, \textsc{\#seniorleadership}, \textsc{\#teamlead}, \textsc{\#teamleader}, \textsc{\#teamleadership}, \textsc{\#teamleading}, \textsc{\#techlead},  \textsc{\#techleader}, \textsc{\#technicalleadership}. 
~The combined searches resulted in a total set of 1,815 retrieved articles.

After collection, the initial set of 1,815 articles underwent a two-phase filtering process to select the most relevant ones. 
In the first phase, one researcher performed a preliminary data cleaning by applying three automatic and simple exclusion criteria: EC1) Removal of duplicate articles (resulting in 53 exclusions); EC2) Exclusion of articles not written in the English language (57 exclusions); and EC3) Exclusion of articles with no audience engagement (reactions and comments) (1,214 exclusions). We applied EC3 to collect articles contained reflections that resonated with or were acknowledged by the practitioner community.

In the second phase, 491 articles were independently analyzed by two researchers, who applied four exclusion criteria based on reading each article. 
To ensure the reliability of the selection process, the researchers initially performed a calibration pilot with 10 articles, refining the exclusion criteria and establishing a shared understanding of the boundaries. Subsequently, during the full analysis of the 491 articles, consensus meetings were held after every batch of 50 articles to discuss borderline cases and maintain calibration throughout the entire process. 
Classification disagreements were resolved by a third senior researcher with extensive experience in qualitative studies in SE. 
The exclusion criteria were: EC4) Articles that only superficially mentioned the \textit{leadership} term or tag without elaborating on the topic (resulting in 216 exclusions); EC5) Articles that addressed leadership but lacked depth or relevance to our research goal (108 exclusions); EC6) Articles from authors whose profiles lacked a biography, a stated job role, or links to external professional networks (e.g., GitHub, LinkedIn), which prevented us from confirming their status as software practitioners and extracting minimum data to characterize the authors (6 exclusions);
and EC7) Articles not contextualized within the SE domain (42 exclusions). 
\textbf{This process resulted in a final and curated selection of 119 articles.} 
The inter-rater agreement for the second phase, measured by Cohen's Kappa coefficient~\cite{McHugh2012}, was 0.879.

\vspace{-0.1cm}

\subsection{Data Analysis}

We define the unit of analysis as the individual article published on Dev.to. The unit of observation corresponds to the specific text segments (sentences or paragraphs) within these articles that describe a leadership behavior or practice. For the analysis of the 119 selected articles, we applied a content analysis approach~\cite{Mayring_2000} to identify recommended and discouraged practices. It is important to note that we adopted a broad definition of ``practice'' to stay true to the practitioners' discourse. In this study, the term encompasses not only specific actionable tasks (e.g., ``1-on-1 Meetings'') but also behavioral competencies and high-level goals (e.g., ``Fostering Team Health'') that practitioners describe as essential duties of a leader. \looseness=-1

The analysis process began with a pilot study to calibrate the data extraction. 
Three articles were randomly selected, and six researchers independently analyzed them to identify and collect the described leadership practices. 
The individual findings were then discussed in a consensus meeting to align understanding and refine the extraction method. 
To avoid bias, the three articles used in the pilot were removed from the data set, resulting in \textbf{116 articles} for the main analysis. 
The selected articles were published on Dev.to from 2017 to 2024, with a higher incidence in 2020 and 2021 (22 articles), and an average length of 1,318 characters.
\looseness=-1

Regarding the analysis of the 116 articles, the researchers were organized into pairs. 
Each pair worked independently to extract text segments describing recommended or discourage leadership practices, without using pre-established codes \cite{strauss1998basics}.
This approach was adopted to respect the diverse nature of the content from the Dev.to.
Throughout the process, the pairs held periodic consensus meetings to discuss and calibrate their findings, ensuring consistency in the extraction. In the end, we extracted 893 text segments divided into two subsets, one with recommended practices segments and another with discouraged ones.

To characterize the practitioners, we manually collected demographic information (such as country, education, job role, and years of experience) by examining each author's Dev.to profile biography or reviewing any externally networks (e.g., LinkedIn, GitHub, X, or personal portfolios) provided by the authors themselves.

\subsubsection{Qualitative Coding} \label{categorization}
Considering the two subsets, one researcher performed manual open coding, resulting in 1,761 initial codes. These codes are textual segments prioritizing the practitioners' own terminology to preserve the authenticity of their discourse before abstracting them into unified practices \cite{rogers2018coding}.
This researcher has experience conducting qualitative studies on human factors in SE and research on leadership within software teams. Subsequently, a senior researcher specializing in qualitative studies on human factors in SE reviewed all extracted codes. Any divergences were resolved through a consensus meeting between the researchers. To exemplify this process, let us consider the following excerpt: ``\ul{\textit{Lead by example. Irrespective of role, everyone has the ability to demonstrate leadership skills and abilities by doing something practical. Demonstrating leadership may involve taking on something complex to support other team members, defusing a conflict, making difficult decisions, having tough conversations etc}}'' (A126276). We identified the initial codes: ``\textit{leading by example},'' ``\textit{defusing conflicts},'' ``\textit{making difficult decisions},'' ``\textit{having tough conversations}'' and ``\textit{taking on complex tasks to support the team}.'' In total, we retained 1,209 coded segments (mentions) across the entire dataset. 
\looseness=-1

By analyzing the codes in each subset, we identified semantically similar codes, which allowed us to unify them. The code unification process was laborious. For example, the initial codes ``\textit{leading by example},'' ``\textit{leading by action},'' and ``\textit{taking responsibilities}'' were unified into ``\textit{showing the way through example},'' while the initial codes ``\textit{defusing conflicts}'' and ``\textit{resolving conflicts}'' were unified into ``\textit{handling conflicts and low performance}.'' 
The process was conducted and reviewed by the same researchers who participated in the initial coding process. They repeatedly revisited the original text segments to disambiguate terms with identical initial codes but different meanings across contexts and, kept distinct practices (e.g., ``Managing 1-on-1s'' vs. ``Managing Technical Meetings'') separate when practitioners explicitly distinguished their purposes and mechanisms, avoiding premature merging that could conflate analytically distinct leadership behaviors.
After unifying the codes, we aimed for a set of unified codes in each list, along with their number of mentions. \looseness=-1

We assessed the final list of unified codes to determine whether theoretical saturation had been reached. Saturation was achieved at the 101st article, as no new codes were identified afterward.

By noticing that many codes in the same subset (recommended or discouraged practices) were related to
each other, we continued the application of open coding to group them into two distinct perspectives \cite{strauss1998basics}. 
The first perspective allowed us to identify the \textbf{\textit{nature of the practices}}, i.e., whether a (recommended or discouraged) practice is \textit{technical} or \textit{managerial}. Technical practices are related to actions or behaviors of the software development process, such as \textbf{Managing and Participating in Code Reviews}, while managerial practices are associated with actions or behaviors of project management, for example \textbf{Learning From One's Own Mistakes}. \looseness=-1

The second perspective allowed us to group the \textbf{\textit{recommended and discouraged practices into categories}}. These categories emerged from the data, revealing distinct areas of concern in leadership. In the following, we briefly present them. The category \textit{People Management \& Development} groups practices focused on the growth, well-being, and empowerment of the team (e.g., \textbf{Fostering a Learning Culture}), while \textit{Strategic Vision} encompasses practices related to direction, long-term planning, and alignment with business objectives (e.g., \textbf{Making Strategic Decisions}). 
The \textit{Processes \& Execution} category has day-to-day operational practices to ensure work flows efficiently and with high quality (e.g., \textbf{Removing Blockers}), and \textit{Communication \& Articulation} includes practices for interacting with the team, stakeholders, and management (e.g., \textbf{Articulating Technical Conversations}). Lastly, \textit{Professional \& Personal Growth} groups practices related to intrinsic qualities adopted by the leader (e.g., \textbf{Learning Continuously}).

To group practices into categories, we based our approach on the primary object of the leadership action (e.g., individual, team, task, or organization) and its temporal focus (e.g., daily execution versus long-term strategy). We then iteratively discussed borderline cases, i.e., instances in which a practice could fit into multiple categories. For example, technical practices such as \textbf{Designing and Overseeing Software Architecture} were assigned to \textit{Processes \& Execution} rather than \textit{Strategic Vision} because practitioners predominantly framed them as operational, day-to-day oversight activities rather than long-term business goals.

\begin{figure*}
    \centering
    \includegraphics[scale=0.36]{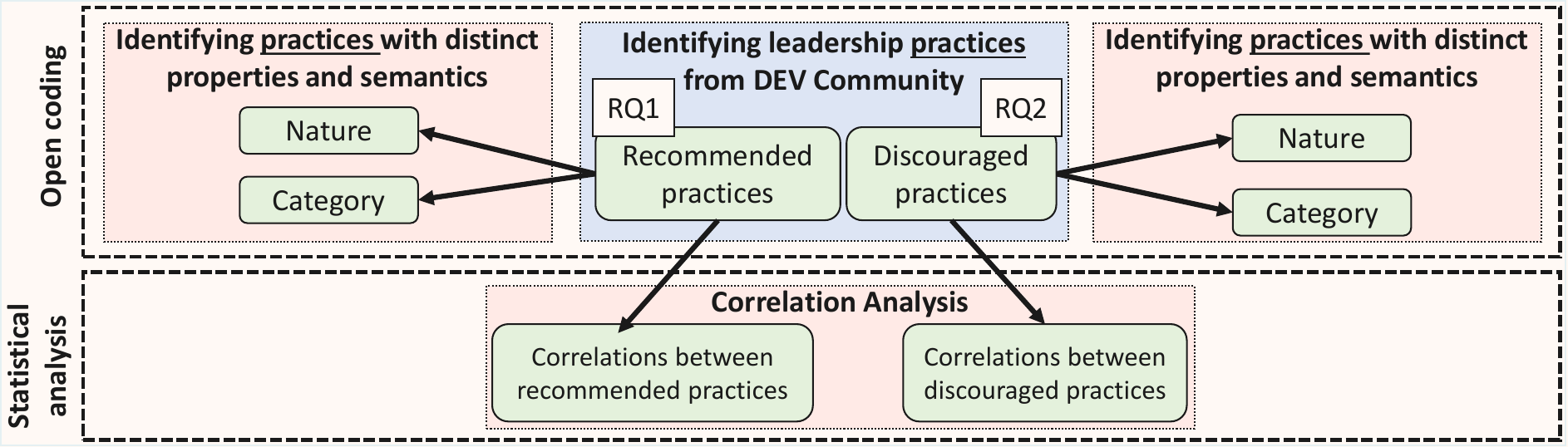}
    \caption{Coding process: Distinct properties and dimensions. Adapted from~\cite{freire2023JSS}.}
    \label{fig:codingprocess}
    \vspace{-0.4cm}
\end{figure*}

The grouping process (from both perspectives) was conducted and reviewed by the researchers who conducted the code unification. Figure~\ref{fig:codingprocess} summarizes the coding process and the perspectives used to organize the codes.
The complete dataset and detailed coding schemes are available in our replication package \cite{repo}.

\subsubsection{Correlation between practices}
We found that several articles contained more than one recommended or discouraged practice, allowing us to investigate the correlations between these practices. We used a correlation matrix and interpreted its values according to Hopkins’s categorization~\cite{hopkins2016}: (i) almost perfect (0.9 to 1.0 or -0.9 to -1.0), (ii) very large (0.7 to 0.9 or -0.7 to -0.9), (iii) large (0.5 to 0.7 or -0.5 to -0.7), (iv) moderate (0.3 to 0.5 or -0.3 to -0.5), (v) small (0.1 to 0.3 or -0.1 to -0.3), and (vi) insubstantial (less than 0.1 and greater than -0.1). We also applied Kendall’s correlation test, which is appropriate for categorical data and does not assume normality, conditions that apply to our data set. \looseness=-1

\section{RESULTS} \label{sec:results}

By analyzing the 116 articles, we identified 90 distinct author profiles, 96.7\% being individual practitioners and 3.3\% belonging to corporate accounts. Moreover, 90\% of the authors included a biography and provided at least one external contact channel (\textit{X}, GitHub, or personal website). 
Most authors are from the United States (35.6\%), followed by the United Kingdom, Canada, and Germany (5.6\% each), and Australia and Brazil (3.3\% each). We also identified authors from Croatia, India, Mexico, the Netherlands, Portugal, South Africa, Spain, Finland, Iraq, Israel, Poland, Romania, and Singapore. Regarding educational background, 21.2\% reported a degree in IT, 2.2\% in Physics, and 2.2\% in Humanities, while 74.4\% did not report any educational information.
Lastly, 11.1\% of the authors hold a Bachelor's degree, 7.8\% a Master's degree, 1.1\% a Doctoral degree, 1.1\% are undergraduates, and 1.1\% are graduate students. However, 77.8\% did not specify their education level. The average professional experience was approximately 8.2 years.
The following subsections present the results per RQs.

\subsection{RQ1: Recommended practices for leaders} \label{RQ1}

We identified 1,113 mentions of recommend leadership practices in the analyzed articles. These mentions were normalized into \textbf{71 distinct practices} as summarized in Figure ~\ref{fig:ConceptualMap}. 

\begin{table}[]
\caption{Top ten recommended practices for leaders in software development projects}
\resizebox{0.97\columnwidth}{!}{%
\begin{tabular}{llll}
\hline
\textbf{NO} & \textbf{Leadership Practices} & \textbf{\#CP} & \textbf{\%PCPA}  \\ \hline
1st & Cultivating \&  Practicing Interpersonal Skills  & 55   & 47.4\% \\
2nd & Managing \&  Delegating Team Work  & 46   & 39.6\% \\
3rd & Practicing \&  Developing Managerial Skills     & 41    & 35.3\% \\
4th & Coaching \&  Mentoring Team Members    & 38   & 32.7\% \\
5th & Fostering Collaboration \&  Empathy    & 33    & 28.4\% \\
6th & Building \&  Sustaining Team Culture   & 33    & 28.4\% \\
7th & Having \&  Developing a Good Communication Skill   & 30    & 25.8\% \\
8th & Supporting Team Members  & 29    & 25.0\% \\
9th & Promoting Team Motivation \&  Engagement  & 28    & 24.1\% \\  
10th & Practicing Continuous Learning \&  Growth     & 27    & 23.2\% \\ \hline
\multicolumn{4}{p{0.95\columnwidth}}{\scriptsize
\textbf{Caption:} \#CP: Count of leadership practice mentions. 
\%PCPA: Percentage of \#CP in relation to all articles (N=116).
} \\ \hline
\end{tabular}
}
\label{tab:LeadershipPractices}
\end{table}

By analyzing Table~\ref{tab:LeadershipPractices} (which presents the ten most mentioned recommended practices), we note a predominance of managerial and interpersonal practices. The most frequently mentioned practice, \textbf{Cultivating \& Practicing Interpersonal Skills}, appeared in nearly half of the analyzed articles (47.4\%), emphasizing the importance of soft skills for leaders. This trend is reinforced by the subsequent practices, which highlight team management (\textbf{Managing \& Delegating Team Work}), people development (\textbf{Coaching \& Mentoring Team Members}), and the creation of a collaborative environment (\textbf{Fostering Collaboration \& Empathy} and \textbf{Building \& Sustaining Team Culture}). From the practitioners' viewpoint, a leader's ability to orchestrate teamwork and cultivate a psychologically safe and motivating environment is important. \looseness=-2

In the context of this work, \textbf{Cultivating \& Practicing Interpersonal Skills} refers to interpersonal skills that practitioners value in a leader, as illustrated in ``\ul{\textit{If it can be acceptable to be a bit too blunt as a peer developer, as a lead you have to be much more careful. I immediately recognized that I had to work on my people and communication skills}}'' (A534004). 
\textbf{Managing \& Delegating Team Work} represents managerial practices focused on work organization, as shown in ``\ul{\textit{As a lead/manager, that focus became delegating the fun stuff so others can do the work, looking at the big picture, facilitating the conversation for others to find solutions, and ensuring that they all level up}}'' (A1556597). 
\textbf{Practicing \& Developing Managerial Skills} covers the development of the leader's own management competencies, as in ``\ul{\textit{Keep developing your leadership skills: The best way to boost your leadership skills is via on-the-job training, with regular feedback and coaching to help you to continually hone your skills}}'' (A126276).

\textbf{Coaching \& Mentoring Team Members} represents actions for team guidance and development, as in ``\ul{\textit{Be your team's biggest advocate. Get to know your team in a professional setting, through regular 1:1s. Get to know their interests and how you can help them grow their career. Be a coach and a mentor. Recognize the value that each member brings to the team and give positive feedback generously}}'' (A135508). 
\textbf{Fostering Collaboration \& Empathy} consolidates practices that promote collaboration and empathy, as shown in ``\ul{\textit{At least once per sprint, have a knowledge-sharing meetings just for the developers where everyone discusses the work they did, and the challenges they faced}}'' (A73231). 
\textbf{Building \& Sustaining Team Culture} means actions that build and maintain team culture, for example ``\ul{\textit{Building a strong team culture increases the whole team’s productivity and efficiency. Having the necessary skills and knowledge is important, but partaking in building team culture will bring cohesiveness to the next level.}}'' (A1692154). 
\textbf{Having \& Developing a Good Communication Skill} refers to effective communication practices, such as ``\ul{\textit{You can achieve almost anything by teamwork. Listening is a vital characteristic of a leader. Effectively communicating your thoughts is essential, as well. How you interact can positively and negatively affect the relationships you have in your work (and personal life too)}}'' (A295707).

Lastly, \textbf{Supporting Team Members} considers direct support practices for the team, as evidenced in ``\ul{\textit{Additionally, you’re there to provide continuous support and encouragement to your team (\dots)}}'' (A1692154). 
\textbf{Promoting Team Motivation \& Engagement} reflects practices to boost team engagement, as shown in ``\ul{\textit{Apart from team agreements, bear in mind that a team is made of people, and a team lead's most important task is to motivate them, making them happy to work on the project. In order to achieve this it is important to get to know them, their sweet and weak spots both from a tech and non tech perspective. The idea is to know how to empower them and make them grow defining a path to follow}}'' (A845150). 
And \textbf{Practicing Continuous Learning \& Growth}: describes practices focused on the leader's self-development, as described in ``\ul{\textit{Remember, the key to being a better manager lies in continuous learning and adapting to the evolving needs of your team}}'' (A1771644).

\begin{tcolorbox}[colback=gray!10, colframe=black, width=\linewidth]
\textbf{Finding \#1}: 
From the articles, leadership in software projects is predominantly characterized by managerial practices. Interpersonal, team management, and people development skills are perceived as more central to a leader's success than their isolated technical expertise.
\end{tcolorbox}

\subsubsection{Technical recommended practices}
Looking at Table~\ref{tab:LeadershipPractices}, it becomes evident that all the identified recommended practices are managerial in nature. However, we also identified technical recommended practices in our data set. In total, 87.4\% (55 practices) were classified as managerial, whereas 12.6\% (16 practices) required formal technical knowledge for their execution. 

The most mentioned technical recommended practice, \textbf{Technical Expertise \& Domain Knowledge} (18 mentions; 15.5\% of mentions total), acts as the foundation, suggesting that credibility and a deep understanding of the product and technology are essential. This practice reflects technical practices necessary for the professional to exercise the leadership function, as shown in ``\ul{\textit{Staying technical is essential. In software, discreteness and rational thought are paramount. The language and discipline of engineering ignite the performance of others. Managers know their limits and stay higher-ordered. Whereas the team has latitude with the craft, EMs develop a toolbox to elevate that craft further}}'' (A252945). \looseness=-1

\textbf{Managing \& Participating in Code Reviews} (14 mentions; 12\% of mentions total) refers to actively participating in the code review process, as demonstrated in ``\ul{\textit{Read code and participate in code reviews. You may not be able to contribute to sprint cycles regularly anymore. Keep in the know by reading code that is being written and merged to production every sprint}}'' (A135508). 
\textbf{Ensuring Code Quality \& Technical Standards} (12 mentions; 10.3\% of mentions total) guarantee the software quality process, as shown in ``\ul{\textit{You are responsible for ensuring that teams satisfy high-quality standards and that best practices are implemented while delivering all projects and features on time}}'' (A73231). Together these practices emphasize technical oversight and enablement, positioning the leader as a guardian of the team's technical quality. 

Furthermore, practices like \textbf{Practicing Effective Documentation} (12 mentions; 10.3\% mentions total) and \textbf{Managing Requirements \& Product Development} (11 mentions; 9.4\% mentions total) connect the leader's technical role directly to process quality and product strategy. 
\textbf{Practicing Effective Documentation} consolidates actions related to project documentation, as we can see ``\ul{\textit{Adequate documentation supports both the development team and stakeholders, and it’s an essential part of every software project. Doing this correctly will help communicate more effectively and prevent future mistakes}}'' (A1692154). 
And \textbf{Managing Requirements \& Product Development} points to practices ranging from requirements gathering to product domain, for example ``\ul{\textit{The Team Lead must have a sufficient understanding of the requirements of the new feature and the desired outcome. In my experience, it is often the case that no one has a complete picture of the proposed solution (or even the problem!) before the development team starts planning how to build the new feature or even before they ship the first iteration}}'' (A207603).

\begin{tcolorbox}[colback=gray!10, colframe=black, width=\linewidth]
\textbf{Finding \#2}: 
Although technical practices were less frequent in the analyzed articles, practitioners valued leaders who guide strategic decisions, ensure technical quality (via code reviews and standards), and demonstrate strong product and domain knowledge.
\end{tcolorbox}

\subsubsection{Categories of recommended practices}

We grouped the 71 distinct recommended practices into five categories. Table \ref{tab:Classification} details this distribution. \textit{Processes \& Execution} (21 practices) is the category with the most distinct practices, followed by \textit{People Management \& Development} (18), \textit{Communication \& Articulation} (13), \textit{Strategic Vision} (10), and \textit{Professional \& Personal Growth} (9).

\begin{table}[]
\centering
\caption{Distribution of Recommended Practices by Category and Type (Managerial and Technical)}
\label{tab:Classification}
\resizebox{0.97\columnwidth}{!}{%
\begin{tabular}{p{3cm}p{0.3cm}p{0.3cm}p{0.3cm} p{0.3cm}p{0.3cm}p{0.3cm} p{0.3cm}p{0.3cm}p{0.7cm}}
\hline
\multirow{2}{*}{
\diagbox[width=3cm,height=2.2em]{\textbf{Category}}{\textbf{Type}}} 
& \multicolumn{3}{c}{\makecell{\textbf{Managerial} \\ \textbf{Practices (MP)}}}
& \multicolumn{3}{c}{\makecell{\textbf{Technical} \\ \textbf{Practices (TP)}}}
& \multicolumn{3}{c}{\makecell{\textbf{Total} \\ \textbf{(MP + TP)}}} \\
\cline{2-10}
 & \textbf{\#P} & \textbf{\#M} & \textbf{\%M} & \textbf{\#P} & \textbf{\#M} & \textbf{\%M} & \textbf{\#P} & \textbf{\#M} & \textbf{\%M} \\ \hline
People Management \& Development & 17 & 341 & 30.6\% & 1 & 4 & 0.3\% & \textbf{18} & \textbf{345} & \textbf{31.0\%} \\
Communication \& Articulation & 12 & 171 & 15.3\% & 1 & 9 & 0.8\% & \textbf{13} & \textbf{180} & \textbf{16.2\%} \\
Processes \& Execution & 10 & 175 & 15.7\% & 11 & 90 & 8.0\% & \textbf{21} & \textbf{265} & \textbf{23.8\%} \\
Professional \& Personal Growth & 8 & 183 & 16.4\% & 1 & 18 & 1.6\% & \textbf{9} & \textbf{201} & \textbf{18.1\%} \\
Strategic Vision & 8 & 103 & 9.25\% & 2 & 19 & 1.7\% & \textbf{10} & \textbf{122} & \textbf{11.0\%} \\ \hline
\textbf{Total} & \textbf{55} & \textbf{973} & \textbf{87.4\%} & \textbf{16} & \textbf{140} & \textbf{12.6\%} & \textbf{71} & \textbf{1113} & \textbf{100.0\%} \\ \hline
\multicolumn{10}{p{0.95\columnwidth}}{\scriptsize
\textbf{Legend:} 
\#P: Number of distinct practices (out of 71 total). 
\#M: Total number of mentions (out of 1,113 total). 
\%M: Percentage of mentions relative to the total (1,113).
} \\ \hline
\end{tabular}
}
\end{table}

Table~\ref{tab:Classification} also shows the distribution of managerial and technical practices in each category. 
While the \textit{Processes \& Execution} category has the highest number of distinct practices, \textit{People Management \& Development} is the most dominant category in terms of practitioner focus (31.0\% of all mentions). 
These categories are almost entirely composed of managerial practices. 
In contrast, \textit{Processes \& Execution} is the only category where technical practices have a significant weight (8.0\% of the total), reinforcing the idea that a leader's technical contributions are primarily focused on process and quality oversight. The other categories, \textit{Professional \& Personal Growth}, \textit{Communication \& Articulation}, and \textit{Strategic Vision}, are all overwhelmingly composed of managerial practices.

\begin{tcolorbox}[colback=gray!10, colframe=black, width=\linewidth]
\textbf{Finding \#3}:
As \textit{People Management \& Development}, \textit{Professional \& Personal Growth}, \textit{Communication \& Articulation}, and \textit{Strategic Vision} are predominantly managerial, and \textit{Processes \& Execution} captures the main technical component, leadership primarily focuses on people and strategy, with technical engagement centered on process, quality, and execution. \looseness=-1
\end{tcolorbox}

\subsubsection{Correlation between the recommended practices}
\begin{figure}
   \centering
   \includegraphics[trim=1cm 2.8cm 2.8cm 2.5cm, clip, scale=0.42]{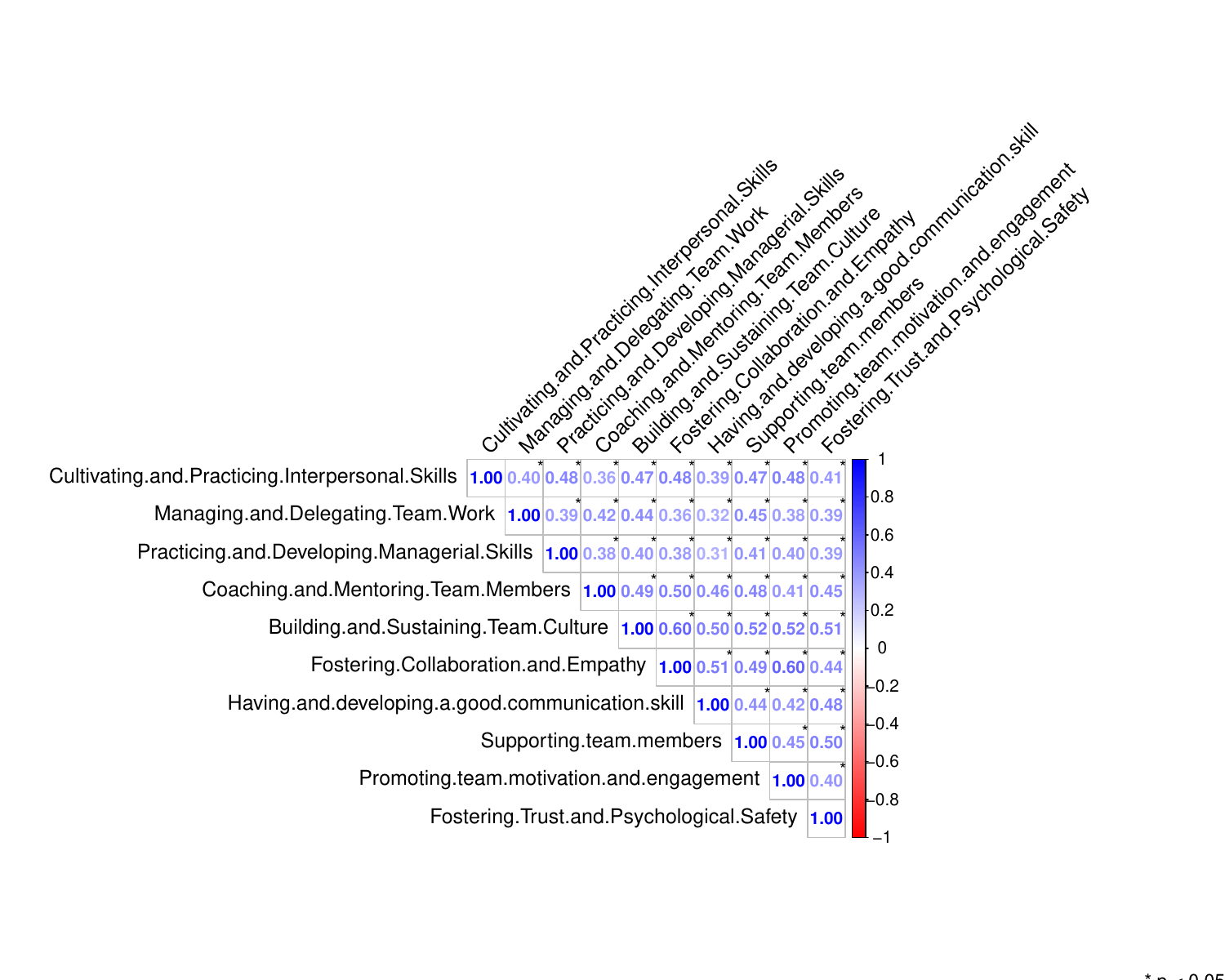}
    \caption{Correlation matrix between the top ten recommended practices. The asterisk (*) indicates a coefficient with statistical significance.}
    \label{fig:postivePracticesCorrelation}
\end{figure}

We analyzed 103 articles containing multiple recommended practices to explore how they co-occur. 
Kendall’s \textit{Tau} coefficients were calculated for each pair among the top 10 recommended practices listed in Table~\ref{tab:LeadershipPractices}. 
Figure~\ref{fig:postivePracticesCorrelation} shows the correlation matrix, where rows and columns represent recommended practices and coefficient strength follows Hopkins’ categorization~\cite{hopkins2016}. 
Colors indicate positive (blue) and negative (red) correlation, and statistically significant pairs are marked with an asterisk (*). 
The full set of correlation values and related statistics is available in our replication package~\cite{repo}.

As shown in Figure~\ref{fig:postivePracticesCorrelation}, Kendall’s \textit{Tau} coefficients range from 0.31 to 0.60, indicating moderate to large effects. Statistical testing confirmed that all top 10 recommended practices are significantly correlated. These results suggest that when one recommended practice is discussed, other related recommended practices are likely to be present as well. 
For example, when the \textbf{Building \& Sustaining Team Culture} practice is discussed, \textbf{Fostering Collaboration \& Empathy} often appears in conjunction, indicating that practitioners frequently mention these behaviors together, often presenting them as a clustered set of advice.

\begin{tcolorbox}[colback=gray!10, colframe=black, width=\linewidth]
\textbf{Finding \#4}: The ten most frequently mentioned recommended practices show statistically significant correlations, suggesting that these recommended practices tend to be discussed together, forming common themes of leadership advice rather than isolated points.
\end{tcolorbox}

\subsection{RQ2: Discouraged practices for leaders}

We identified 96 mentions of leadership practices that should be avoided, which were normalized into \textbf{32 distinct practices}, as shown in Figure~\ref{fig:ConceptualMap}. Table \ref{tab:AvoiedLeadershipPractices} presents the ten most mentioned discouraged practices.

\begin{table}[]
\caption{Top ten discouraged practices for leaders in software development projects}
\resizebox{0.97\columnwidth}{!}{%
\begin{tabular}{llll}
\hline
\textbf{NO} & \textbf{Discouraged Leadership Practices} & \textbf{\#CP} & \textbf{\%PCPA}  \\ \hline
1st & Micromanagement  & 9   & 7.7\% \\
2nd & Counterproductive Work Patterns  & 8   & 6.8\% \\
3rd & Counterproductive Communication Styles     & 7   & 6.0\% \\
4th & Unhealthy Work Practices    & 6   & 5.1\% \\
5th & Excessive Self-Pressure    & 6   & 5.1\% \\
6th & Arrogance and Ego   & 6    & 5.1\% \\
7th & Unethical and Unfair Treatment   & 5    & 4.3\% \\
8th & Authoritarianism & 5    & 4.3\% \\
9th & Fixed Mindset & 5    & 4.3\% \\  
10th & Unhelpful Advice & 3    & 2.5\% \\ \hline
\multicolumn{4}{p{0.95\columnwidth}}{\scriptsize
\textbf{Caption:} \#CP: Count of discouraged leadership practice mentions. 
\%PCPA: Percentage of \#CP in relation to all articles (N=116).
} \\ \hline
\end{tabular}
}
\label{tab:AvoiedLeadershipPractices}
\end{table}

Looking at Table \ref{tab:AvoiedLeadershipPractices}, one can see that the most mentioned discouraged practice is \textbf{Micromanagement} (7.7\%). 
It refers to actions of excessive control over people and tasks, as illustrated in the excerpt: ``\ul{\textit{This is the stereotypical development practice in a waterfall shop. With this mindset: Decisions are made at the top and given to the dev teams for implementation; Managers assign tasks to developers who are treated as kids that need to be reminded not to slack off, also known as command-and-control}}'' (A29008). 
Next comes the practices of \textbf{Counterproductive Work Patterns} (6.8\%) and \textbf{Counterproductive Communication Styles} (6.0\%). The first points to practices that negatively impact team productivity, such as constant context switching: ``\ul{\textit{Focus factor (context switching) Even when a team member has a healthy WIP, they can still be affected by what I call the ``focus factor''. If you’re working on several different types of stories in a two week sprint your focus can be spread too thin. It stops you from going deep into customer pain points and ideal user experience}}'' (A453703), while the latter indicates the need to avoid aggressive or unnecessary communication styles, especially in times of crisis: ``\ul{\textit{Avoid passive-aggressive comments. Whether you know it or not, your Psychological safety skill is to test when shit hits the fan. You have time to think when things are good, but when they are not, you don't. This is when you need to be careful and avoid any passive-aggressive comments. Emotions would run high during this time, not just for you but for your entire team, so make sure to keep your emotions in check}}'' (A1239086). \looseness=-1

The discouraged practices of \textbf{Unhealthy Work Practices}, \textbf{Excessive Self-pressure}, and \textbf{Arrogance and Ego} had each one the same percentage of mentions (5.1\%). \textbf{Unhealthy Work Practices} consolidates practices detrimental to the team's long-term health, such as work overload: ``\ul{\textit{However, overloading employees with work is counterproductive in the long run. Smart leaders understand this and prioritize a healthy work schedule for their team members}}'' (A1391206). 
\textbf{Excessive Self-pressure}, in its turn, refers to avoiding excessive pressure on oneself: ``\ul{\textit{One day you will think you know everything, but you may feel the opposite the next, so don't be too hard on yourself and keep learning}}'' (A1391395). 
\textbf{Arrogance and Ego} contains practices that demonstrate arrogance from the leader, suggesting an emphasis on empowering the team instead: ``\ul{\textit{If you think that you are the most important person in the room, it’s time to flip your perspective. Instead of being part of every discussion, empower the team to run without you}}'' (A24182).

Lastly, the discouraged practices of \textbf{Unethical and Unfair Treatment}, \textbf{Authoritarianism}, and \textbf{Fixed Mindset} had each one 4.3\% of mentions, while \textbf{Unhelpful Advice} had 2.5\%. 
\textbf{Unethical and Unfair Treatment} addresses ethically inappropriate situations, such as the use of gender-biased language: ``\ul{\textit{Language matters. Make sure you avoid gender bias from your content as much as possible. I recommend you run your content through an analyser such as Gender Decoder}}'' (A678843). 
\textbf{Authoritarianism} highlights that leadership in software development is often perceived as less prescriptive than in other fields, focusing on support and mentorship: ``\ul{\textit{However, leadership in software development can be very different then other fields. It's less authoritative and prescriptive in other fields, and is focused on providing support and mentor-ship to other developers. That being said, classical leadership skills are always useful}}'' (A344928). 
Finally, \textbf{Fixed Mindset} contextualizes practices where the leader makes assumptions before discussing with the team: ``\ul{\textit{is to avoid making assumptions and simply ask for an explanation during one of your regular check-ins. The developer probably just encountered issues that you were not aware of, and it won’t become a repeating issue}}'' (A73231), and \textbf{Avoiding Unhelpful Advice} indicates that leaders are encouraged to resist providing advice that does not add value: ``\ul{\textit{You must resist the urge to dive into advice-giving mode}}'' (A126276).

\begin{tcolorbox}[colback=gray!10, colframe=black, width=\linewidth]
\textbf{Finding \#5}: 
Across the articles, discouraged practices are predominantly interpersonal, cultural, and managerial rather than technical. \textbf{Micromanagement} is the most mentioned practice, followed by behaviors that undermine team autonomy and culture, including counterproductive work and unhealthy patterns, and harmful communication.
\end{tcolorbox}

\subsubsection{Technical Discouraged Practices}
We found three technical discouraged practices in the data set. 
The discouraged technical practices represent less than 2\% of all mentions, reinforcing that leadership failures perceived by practitioners are primarily managerial and interpersonal. It is noteworthy that the practice of \textbf{Unilateral Architectural Decisions} (2 mentions; 1.7\% of mentions total) does not criticize the technical decision itself, but rather how it is made. 
Its unilateral nature overlaps with the previously discussed themes of authoritarianism and lack of collaboration, as illustrated in: ``\ul{\textit{the definition of the architecture and implementation of parts of the software system are your responsibility. But that doesn't mean that you should define it on your own. It's a good idea to build something considering the input of your team. It's easier for developers to build something if they contributed to the solution actively}}'' (A763053). 
The other practices (1 mention each; 0.8\% of mentions each) focus on technical errors and the creation of bottlenecks that impact team execution. \textbf{Technical Missteps} is exemplified by the need to prevent over-engineering (``\ul{\textit{the Team Lead’s job here is to make sure that developers don’t try to solve every unexpected problem, overengineer, and delay the project}}'' (A207603)), while \textbf{Cross-Team Data Bottlenecks} points to the importance of not depending on other teams for essential data (``\ul{\textit{Data informs everything. Do it yourself (or delegate within your team). Learn SQL and Statistics 101. Do not depend on other teams. Data is your job}}'' (A126394)).

\begin{tcolorbox}[colback=gray!10, colframe=black, width=\linewidth]
\textbf{Finding \#6}:
Technical discouraged practices are largely absent from practitioners’ concerns about poor leadership. When mentioned, they involve specific execution and decision-making issues, such as unilateral architectural decisions, data bottlenecks, and tolerance of technical errors or over-engineering.
\end{tcolorbox}

\subsubsection{Categories of Discouraged Practices}

We grouped the 32 discouraged practices into five categories. Table \ref{tab:AvoidedClassification} details this distribution.
The categories with the most distinct practices were \textit{Professional \& Personal Growth} (10 practices) and \textit{Processes \& Execution} (7 practices), followed by \textit{People management \& Development} (6 practices), \textit{Communication \& Articulation} (5 practices), and \textit{Strategic Vision} (1 practices). Moreover, the categories with the highest number of mentions are \textit{People Management \& Development} (30.2\%) and \textit{Professional \& Personal Growth} (28.1\%), indicating that the main failures perceived by practitioners are poor people management and inappropriate personal postures (such as arrogance or a fixed mindset). \looseness=-1

\begin{table}[]
\centering
\caption{Distribution of Discouraged Practices by Category and Type (Managerial vs. Technical)}
\label{tab:AvoidedClassification}
\resizebox{.97\columnwidth}{!}{%
\begin{tabular}{p{3cm}p{0.3cm}p{0.3cm}p{0.3cm} p{0.3cm}p{0.3cm}p{0.3cm} p{0.3cm}p{0.3cm}p{0.7cm}}
\hline
\multirow{2}{*}{
\diagbox[width=3cm,height=2.2em]{\textbf{Category}}{\textbf{Type}}} 
& \multicolumn{3}{c}{\makecell{\textbf{Managerial} \\ \textbf{Practices (MP)}}}
& \multicolumn{3}{c}{\makecell{\textbf{Technical} \\ \textbf{Practices (TP)}}}
& \multicolumn{3}{c}{\makecell{\textbf{Total} \\ \textbf{(MP + TP)}}} \\
\cline{2-10} 
 & \textbf{\#P} & \textbf{\#M} & \textbf{\%M} & \textbf{\#P} & \textbf{\#M} & \textbf{\%M} & \textbf{\#P} & \textbf{\#M} & \textbf{\%M} \\ \hline
People Management \& Development & 6 & 29 & 30.2\% & 0 & 0 & 0\% & \textbf{6} & \textbf{29} & \textbf{30.2\%} \\
Communication \& Articulation & 5 & 15 & 15.6\% & 0 & 0 & 0\% & \textbf{5} & \textbf{15} & \textbf{15.6\%} \\
Processes \& Execution & 7 & 20 & 20.8\% & 3 & 4 & 4.2\% & \textbf{10} & \textbf{24} & \textbf{25.0\%} \\
Professional \& Personal Growth & 10 & 27 & 28.1\% & 0 & 0 & 0\% & \textbf{10} & \textbf{27} & \textbf{28.1\%} \\
Strategic Vision & 1 & 1 & 1.0\% & 0 & 0 & 0\% & \textbf{1} & \textbf{1} & \textbf{1.0\%} \\ \hline
\textbf{Total} & \textbf{29} & \textbf{92} & \textbf{95.8\%} & \textbf{3} & \textbf{4} & \textbf{4.2\%} & \textbf{32} & \textbf{96} & \textbf{100.0\%} \\ \hline
\multicolumn{10}{p{0.95\columnwidth}}{\scriptsize
\textbf{Legend:}
\#P: Number of distinct practices (out of 32 total).
\#M: Total number of mentions (out of 96 total).
\%M: Percentage of mentions relative to the total (96).
} \\ \hline
\end{tabular}
}
\end{table}

Table \ref{tab:AvoidedClassification} also shows the managerial and technical practices per category. One can see that the \textit{Processes \& Execution} category (25.0\%) is the only one that concentrates the entirety of the technical discouraged practices, demonstrating that perceived technical failures are linked to process execution, rather than to strategy or people management.

\begin{tcolorbox}[colback=gray!10, colframe=black, width=\linewidth]
\textbf{Finding \#7}: 
Discouraged practices are mainly concentrated in personal conduct and people management, with \textit{People Management \& Development} and \textit{Professional \& Personal Growth} as the main sources of friction. In contrast, \textit{Processes \& Execution} includes technical failures, indicating that perceived leadership shortcomings are largely managerial and interpersonal rather than technical.
\end{tcolorbox}

\subsubsection{Correlation between the discouraged practices}
\begin{figure}
   \centering
   \includegraphics[trim=3.5cm 2.8cm 6cm 1.1cm, clip, scale=0.34]{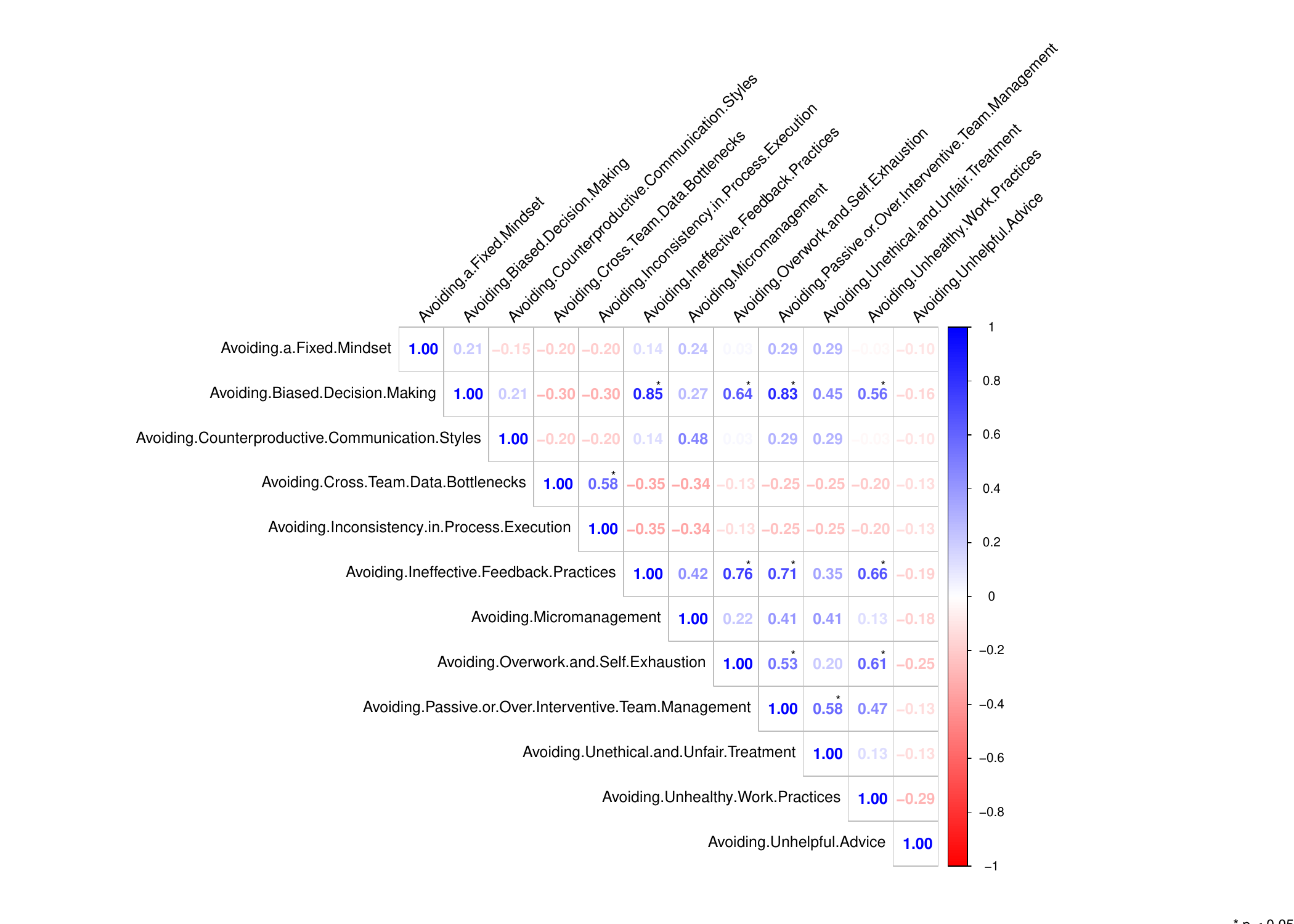}
    \caption{Correlation matrix between discouraged practices. The asterisk (*) indicates a coefficient with statistical significance.\looseness=-1}
    \label{fig:negativePracticesCorrelation}
    \vspace{-0.4cm}
\end{figure}

We analyzed 10 articles containing twelve discouraged practices to explore how they co-occur. Kendall’s \textit{Tau} coefficients were calculated for each pair among these discouraged practices. Figure~\ref{fig:negativePracticesCorrelation} shows the correlation matrix, where rows and columns represent discouraged practices and coefficient strength follows Hopkins’ categorization~\cite{hopkins2016}. 
Colors indicate positive (blue), negative (red), or no (white) correlation, and statistically significant pairs are marked with an asterisk (*). The full set of correlation values and statistics is available in our replication package~\cite{repo}.

As shown in Figure~\ref{fig:negativePracticesCorrelation}, we can see that \textbf{Avoiding Biased Decision Making} is statistically significantly correlated with both \textbf{Avoiding Ineffective Feedback Practices} and \textbf{Avoiding Passive or Over Interventive Team Management}. 
A similar pattern is observed between \textbf{Avoiding Ineffective Feedback Practices}, \textbf{Avoiding Overwork and Self Exhaustion}, \textbf{Avoiding Passive or Over Interventive Team Management}. We also identified other statistically significant correlations with high effect sizes. 

\begin{tcolorbox}[colback=gray!10, colframe=black, width=\linewidth]
\textbf{Finding \#8}:
Statistically significant correlations among discouraged practices show that poor communication, lack of fairness, and excessive control often co-occur, while unbalanced management behaviors further strain individuals and teams. Overall, these results indicate that leadership failures form interconnected behavioral patterns that undermine team effectiveness and well-being.
\end{tcolorbox}
\vspace{-0.4cm}
\begin{figure*}[ht]
    \centering
    \includegraphics[scale=0.33]{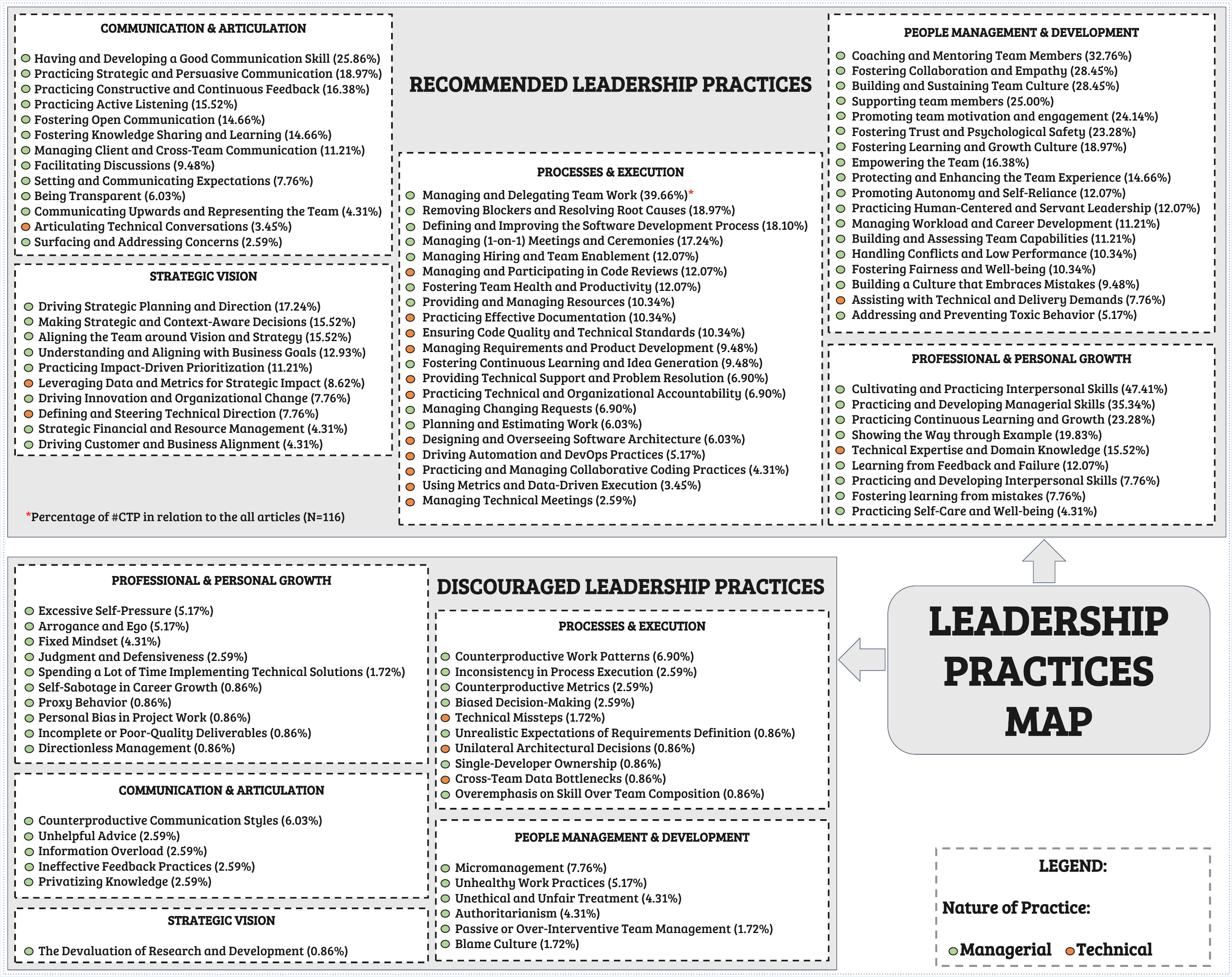}
    \caption{A Conceptual Map of Leadership Practices in Software Engineering}
    \label{fig:ConceptualMap}
\end{figure*}

\section{DISCUSSION} \label{sec:discussion}

\textbf{\textit{Conceptual map.}} To consolidate and structure our findings, we organized the set of recommended and discouraged practices and their categories into a conceptual map, as shown in Figure \ref{fig:ConceptualMap}. The map is based on the structure proposed by~\citet{freire2023JSS}, that serves as a visual structure for formatting and organizing qualitative categories and their frequencies.

The types of practices (recommended or discouraged) are represented by rounded rectangles, whereas the categories are displayed using solid-border rectangles. Each practice includes a percentage indicating its proportion of mentions relative to the total number of analyzed articles (N=116). In addition, each practice is accompanied by a circle that represents its nature, either technical (orange circle) or managerial (green circle).
The map demonstrates that leadership, from the practitioners' perspective, is not a monolithic entity. Instead, it is a complex duality where managerial and interpersonal competence is predominant, yet technical knowledge plays a strategic and well-defined role.

A visual analysis of the conceptual map reveals two central asymmetries. 
First, recommended practices are more diverse (71 items) than discouraged ones (32). While coding granularity may partly explain this, it also suggests practitioners elaborate more on positive role-model behaviors than on specific anti-patterns.
Second, the asymmetry in the technical dimension is stark: while technical practices form a strategic component of the ``Recommended'' side (12.6\% of mentions), they are statistically almost irrelevant on the ``Discouraged'' side (only 4.2\% of mentions). 
This result suggests technical competence is viewed as a bonus for success, but not a primary cause of failure. However, practitioners may attribute technical failures to engineering rather than leadership, effectively excluding them from the “leadership” conversation.

The map shows that \textit{Processes \& Execution} is the only category that mixes  technical and managerial practices in both recommended and discouraged perspectives. This suggests that software leaders are not expected to be primary code contributors but rather to apply technical knowledge to decision-making, quality oversight, and team support. Their main role still remains managerial, supporting teams, fostering personal growth, and promoting strategic vision. 
\\

\noindent \textbf{\textit{Implications.}} The practical implications of this study suggest that organizations may need to reconsider promoting engineers to leadership positions based solely on technical excellence. Our findings demonstrate that the dominant themes in the discourse of software project leaders are predominantly managerial. Beyond its theoretical value, the map offers practical applications for professionals and organizations. It can serve as a competency framework for hiring, enabling recruiters to assess candidates not only on technical fit but also on specific behavioral indicators such as \textit{Fostering Collaboration \& Empathy}. It also can use this taxonomy to conduct skill gap analyses, helping new leaders identify and develop critical interpersonal competencies, such as conflict resolution and empathy, that are often overlooked in technical career paths. Furthermore, it acts as a roadmap for developers aspiring to leadership positions, helping them visualize that their new scope extends beyond code to include management-oriented practices. Finally, mapping discouraged practices offers a checklist of anti-patterns, helping organizations identify and correct toxic behaviors like \textit{Micromanagement} or \textit{Privatizing Knowledge} to prevent harm to the team.

From a research perspective, this study reinforces the value of social media as a valid data source, consistent with prior work on human aspects in SE. Moreover, it fosters a perspective where leadership is understood and enacted through daily actions and practices, rather than being defined solely by hierarchy. Our study offers direct implications for education by supporting instructors in revising curricula for software development courses to explicitly address and cultivate the leadership practices identified in this study. Furthermore, future research could explore pedagogical strategies for effectively teaching and reinforcing these practices to better prepare students for leadership roles in software development projects.
\\

\noindent \textbf{\textit{Related Work.}} \citet{kalliamvakou2017makes} identified key attributes of effective software managers (such as empathy, fairness, and recognition of individuality) showing that managerial effectiveness depends more on supporting teams and facilitating collaboration than on technical expertise. This perspective aligns with our results, where practices such as interpersonal skill development, delegation, and mentoring reflect how these attributes are enacted in everyday work. In another related work, \citet{spiegler2021empirical} examined leadership in agile teams, identifying multiple roles (e.g., facilitator, coach, mediator, and visionary) that initially reside with the Scrum Master but progressively transfer to the team as it matures, depending on factors such as psychological safety, internal support, and role clarity. This dynamic perspective is consistent with our findings on delegation, team support, mentoring, and continuous development as enablers of team autonomy. Finally, \citet{spiegler2021empirical} showed that agile leadership is strongly related to trust, organizational performance, effectiveness, and innovation across contexts.

Overall, prior studies reveal a progression in how leadership has been examined in SE, from managerial competence to agile facilitation and organizational performance. Despite these advances, three gaps remain. First, most research focuses on predefined roles and frameworks, overlooking informal and emergent leadership behaviors. Second, common empirical methods such as surveys and interviews capture structured perspectives but not the spontaneous discourse through which practitioners construct leadership meanings. Third, while meta-analyses associate leadership with performance, they seldom address how practitioners themselves describe or evaluate leadership in daily work. Our study addresses these gaps by adopting a practitioner-centered and discourse-oriented approach, examining how practitioners discuss leadership on social media to uncover how leadership practices are articulated and interpreted in real-world contexts. Furthermore, our findings complement and extend those reported in prior related studies.

\vspace{-0.1cm}
\section{TRUSTWORTHINESS OF THE STUDY} \label{sec:thustworthiness}
To ensure the rigor and trustworthiness of this research, the study was evaluated based on Empirical Standards for SE \cite{ralph2020empirical}. 

To establish \textbf{credibility}, we strengthened the study’s findings by applying strict filtering criteria to select articles with higher engagement and authors with verifiable professional profiles. Furthermore, to mitigate potential biases related to tag incompleteness or inconsistent tagging, we expanded our data retrieval strategy to include 22 related keywords. However, we acknowledge the inherent limitations of our data source. This study maps public discourse and perceptions of leadership rather than actual enacted behaviors. Consequently, the extracted practices reflect what practitioners publicly value and recommend, which may be influenced by self-presentation and social desirability biases. Additionally, our dataset includes only practitioners who choose to share their experiences online, potentially overrepresenting more vocal or community-oriented professionals. Interpretation bias was mitigated through researcher triangulation, including a pilot study, paired analysis, and consensus meetings. \textbf{Confirmability} was further supported by an open coding approach, in which codes emerged directly from the data, and by the peer-review process, ensuring that interpretations were continuously examined and validated.

Regarding \textbf{transferability}, we acknowledge that the findings may not be directly generalizable to all software engineering contexts. Dev.to has a distinct culture, and our results are inherently bounded by this platform. Although it represents a large, global developer community, it does not capture the entire population, particularly the ``silent majority'' who do not engage in online blogging. Therefore, the findings should be interpreted as a characterization of community-validated leadership discourse rather than universally applicable behaviors across all corporate environments.

Finally, the study's \textbf{dependability} was established through a transparent and detailed research protocol, as described in Section~\ref{sec:researchmethod}, with a replication package \cite{repo} that enables an audit of the process.
\section{CONCLUSION} \label{sec:conclusion}

This study aims to identify and characterize leadership practices in SE from the perspective of industry practitioners by analyzing 116 articles from the Dev.to. The contributions of this work are 1) a conceptual map synthesizing recommended and discouraged leadership practices and their corresponding categories, and 2) a comprehensive open repository containing the 116 analyzed articles and detailed (initial and unified) coding, enabling transparency, auditing, and reuse by the research community.

As future work, we plan to extend this study by incorporating additional social media platforms (such as Stack Exchange) to strengthen the reliability and generalizability of the findings and to conduct interviews with software project leaders to assess the practices identified in this paper.

\section*{ACKNOWLEDGMENTS}

This work was supported by the Conselho Nacional de Desenvolvimento Científico e Tecnológico (CNPq), Brazil [Grant 404406/2023-8]; and by the Ceará Foundation for the Support of Scientific and Technological Development (Funcap), Brazil. We used Large Language Models for language refinement, but all analysis and interpretation remain our sole responsibility.

\section*{ARTIFACTS AVAILABILITY}

To ensure transparency and replicability, all data supporting this study are openly accessible via our public repository \cite{repo}.

\bibliographystyle{ACM-Reference-Format}
\balance
\bibliography{reference}


\begin{thebibliography}{45}


\ifx \showCODEN    \undefined \def \showCODEN     #1{\unskip}     \fi
\ifx \showISBNx    \undefined \def \showISBNx     #1{\unskip}     \fi
\ifx \showISBNxiii \undefined \def \showISBNxiii  #1{\unskip}     \fi
\ifx \showISSN     \undefined \def \showISSN      #1{\unskip}     \fi
\ifx \showLCCN     \undefined \def \showLCCN      #1{\unskip}     \fi
\ifx \shownote     \undefined \def \shownote      #1{#1}          \fi
\ifx \showarticletitle \undefined \def \showarticletitle #1{#1}   \fi
\ifx \showURL      \undefined \def \showURL       {\relax}        \fi
\providecommand\bibfield[2]{#2}
\providecommand\bibinfo[2]{#2}
\providecommand\natexlab[1]{#1}
\providecommand\showeprint[2][]{arXiv:#2}

\bibitem[Adanigbo et~al\mbox{.}(2025)]%
        {adanigboconceptual2025}
\bibfield{author}{\bibinfo{person}{Oluwasanmi~Segun Adanigbo}, \bibinfo{person}{Denis Kisina}, \bibinfo{person}{Andrew~Ifesinachi Daraojimba}, \bibinfo{person}{Samuel Owoade}, \bibinfo{person}{Nneka~Adaobi Ochuba}, {and} \bibinfo{person}{Toluwase~Peter Gbenle}.} \bibinfo{year}{2025}\natexlab{}.
\newblock \showarticletitle{A Conceptual Framework for Scaling Technical Leadership and Mentorship in Remote Software Engineering Teams}.
\newblock \bibinfo{journal}{\emph{International Journal of Engineering and Modern Technology (IJEMT)}} \bibinfo{volume}{11}, \bibinfo{number}{4} (\bibinfo{year}{2025}), \bibinfo{pages}{6--12}.
\newblock


\bibitem[Caldiera and Rombach(1994)]%
        {caldiera1994goal}
\bibfield{author}{\bibinfo{person}{Victor R Basili-Gianluigi Caldiera} {and} \bibinfo{person}{H~Dieter Rombach}.} \bibinfo{year}{1994}\natexlab{}.
\newblock \showarticletitle{Goal question metric paradigm}.
\newblock \bibinfo{journal}{\emph{Encyclopedia of software engineering}} \bibinfo{volume}{1}, \bibinfo{number}{528-532} (\bibinfo{year}{1994}), \bibinfo{pages}{6}.
\newblock


\bibitem[Capretz(2014)]%
        {capretz2014bringing}
\bibfield{author}{\bibinfo{person}{Luiz~Fernando Capretz}.} \bibinfo{year}{2014}\natexlab{}.
\newblock \showarticletitle{Bringing the human factor to software engineering}.
\newblock \bibinfo{journal}{\emph{IEEE software}} \bibinfo{volume}{31}, \bibinfo{number}{2} (\bibinfo{year}{2014}), \bibinfo{pages}{104--104}.
\newblock


\bibitem[Cerqueira et~al\mbox{.}(2025)]%
        {Cerqueira2025TOSEM}
\bibfield{author}{\bibinfo{person}{Lidiany Cerqueira}, \bibinfo{person}{Jo\~{a}o~Pedro Bastos}, \bibinfo{person}{Danilo Neves}, \bibinfo{person}{Glauco Carneiro}, \bibinfo{person}{Rodrigo Spinola}, \bibinfo{person}{S\'{a}vio Freire}, \bibinfo{person}{Jose Santos}, {and} \bibinfo{person}{Manoel Mendon\c{c}a}.} \bibinfo{year}{2025}\natexlab{}.
\newblock \showarticletitle{Exploring Empathy in Software Engineering: Insights from a Grey Literature Analysis of Practitioners’ Perspectives}.
\newblock \bibinfo{journal}{\emph{ACM Trans. Softw. Eng. Methodol.}} (\bibinfo{date}{July} \bibinfo{year}{2025}).
\newblock
\showISSN{1049-331X}
\href{https://doi.org/10.1145/3748721}{doi:\nolinkurl{10.1145/3748721}}
\newblock
\shownote{Just Accepted}.


\bibitem[Cerqueira et~al\mbox{.}(2023)]%
        {cerqueira2023thematic}
\bibfield{author}{\bibinfo{person}{Lidiany Cerqueira}, \bibinfo{person}{S{\'a}vio Freire}, \bibinfo{person}{Jo{\~a}o Bastos}, \bibinfo{person}{Rodrigo Sp{\'\i}nola}, \bibinfo{person}{Manoel Mendon{\c{c}}a}, {and} \bibinfo{person}{Jos{\'e} Santos}.} \bibinfo{year}{2023}\natexlab{}.
\newblock \showarticletitle{A thematic synthesis on empathy in software engineering based on the practitioners' perspective}. In \bibinfo{booktitle}{\emph{Proceedings of the XXXVII Brazilian Symposium on Software Engineering}}. \bibinfo{pages}{332--341}.
\newblock


\bibitem[Cerqueira et~al\mbox{.}(2024)]%
        {cerqueira2024IEEESW}
\bibfield{author}{\bibinfo{person}{Lidiany Cerqueira}, \bibinfo{person}{Sávio Freire}, \bibinfo{person}{Danilo~Ferreira Neves}, \bibinfo{person}{João Pedro~Silva Bastos}, \bibinfo{person}{Beatriz Santana}, \bibinfo{person}{Rodrigo Spínola}, \bibinfo{person}{Manoel Mendonça}, {and} \bibinfo{person}{José Amancio~Macedo Santos}.} \bibinfo{year}{2024}\natexlab{}.
\newblock \showarticletitle{Empathy and Its Effects on Software Practitioners’ Well-Being and Mental Health}.
\newblock \bibinfo{journal}{\emph{IEEE Software}} \bibinfo{volume}{41}, \bibinfo{number}{4} (\bibinfo{date}{July} \bibinfo{year}{2024}), \bibinfo{pages}{95--104}.
\newblock
\showISSN{1937-4194}
\href{https://doi.org/10.1109/MS.2024.3377897}{doi:\nolinkurl{10.1109/MS.2024.3377897}}


\bibitem[Coelho et~al\mbox{.}(2024)]%
        {coelho2024estudo}
\bibfield{author}{\bibinfo{person}{Murilo Coelho}, \bibinfo{person}{Allysson~Allex Ara{\'u}jo}, \bibinfo{person}{S{\'a}vio Freire}, {and} \bibinfo{person}{Matheus Paixao}.} \bibinfo{year}{2024}\natexlab{}.
\newblock \showarticletitle{Um estudo comparativo entre a vis{\~a}o de l{\'\i}deres e liderados sobre a import{\^a}ncia de soft skills em desenvolvimento de software}. In \bibinfo{booktitle}{\emph{Workshop sobre Aspectos Sociais, Humanos e Econ{\^o}micos de Software (WASHES)}}. SBC, \bibinfo{pages}{130--140}.
\newblock


\bibitem[Coelho et~al\mbox{.}(2025)]%
        {coelho2025soft}
\bibfield{author}{\bibinfo{person}{Murilo Coelho}, \bibinfo{person}{Allysson~Allex Ara{\'u}jo}, \bibinfo{person}{S{\'a}vio Freire}, {and} \bibinfo{person}{Matheus Paixao}.} \bibinfo{year}{2025}\natexlab{}.
\newblock \showarticletitle{What soft skills are most valued in software projects? A comparative view of leader and non-leader perspectives}.
\newblock \bibinfo{journal}{\emph{iSys-Brazilian Journal of Information Systems}} \bibinfo{volume}{18}, \bibinfo{number}{1} (\bibinfo{year}{2025}), \bibinfo{pages}{9--1}.
\newblock


\bibitem[Coelho et~al\mbox{.}(2026)]%
        {repo}
\bibfield{author}{\bibinfo{person}{Murilo Coelho}, \bibinfo{person}{Denivan Campos}, \bibinfo{person}{Mariana~Maia Bezerra}, \bibinfo{person}{Matheus Paixao}, \bibinfo{person}{Allysson~Allex Araújo}, {and} \bibinfo{person}{Sávio Freire}.} \bibinfo{year}{2026}\natexlab{}.
\newblock \bibinfo{title}{Replication package for the Paper ``What Characterizes a Software Leader? Identifying Leadership Practices from Practitioners’ Social Media''}.
\newblock
\urldef\tempurl%
\url{https://zenodo.org/records/19303421}
\showURL{%
\tempurl}


\bibitem[Faraj and Sambamurthy(2006)]%
        {faraj2006leadership}
\bibfield{author}{\bibinfo{person}{Samer Faraj} {and} \bibinfo{person}{V Sambamurthy}.} \bibinfo{year}{2006}\natexlab{}.
\newblock \showarticletitle{Leadership of information systems development projects}.
\newblock \bibinfo{journal}{\emph{IEEE Transactions on engineering management}} \bibinfo{volume}{53}, \bibinfo{number}{2} (\bibinfo{year}{2006}), \bibinfo{pages}{238--249}.
\newblock


\bibitem[Freire et~al\mbox{.}(2023a)]%
        {freireREFSQ}
\bibfield{author}{\bibinfo{person}{S\'{a}vio Freire}, \bibinfo{person}{Felipe Gomes}, \bibinfo{person}{Larissa Barbosa}, \bibinfo{person}{Thiago~Souto Mendes}, \bibinfo{person}{Galdir Reges}, \bibinfo{person}{Rita S.~P. Maciel}, \bibinfo{person}{Manoel Mendon\c{c}a}, {and} \bibinfo{person}{Rodrigo Sp\'{\i}nola}.} \bibinfo{year}{2023}\natexlab{a}.
\newblock \showarticletitle{Requirements Engineering Issues Experienced by Software Practitioners: A Study on Stack Exchange}. In \bibinfo{booktitle}{\emph{Requirements Engineering: Foundation for Software Quality: 29th International Working Conference, REFSQ 2023, Barcelona, Spain, April 17–20, 2023, Proceedings}} (Barcelona, Spain). \bibinfo{publisher}{Springer-Verlag}, \bibinfo{address}{Berlin, Heidelberg}, \bibinfo{pages}{3–20}.
\newblock
\showISBNx{978-3-031-29785-4}
\href{https://doi.org/10.1007/978-3-031-29786-1_1}{doi:\nolinkurl{10.1007/978-3-031-29786-1_1}}


\bibitem[Freire et~al\mbox{.}(2023b)]%
        {freire2023JSS}
\bibfield{author}{\bibinfo{person}{Sávio Freire}, \bibinfo{person}{Nicolli Rios}, \bibinfo{person}{Boris Pérez}, \bibinfo{person}{Camilo Castellanos}, \bibinfo{person}{Darío Correal}, \bibinfo{person}{Robert Ramač}, \bibinfo{person}{Vladimir Mandić}, \bibinfo{person}{Nebojša Taušan}, \bibinfo{person}{Gustavo López}, \bibinfo{person}{Alexia Pacheco}, \bibinfo{person}{Manoel Mendonça}, \bibinfo{person}{Davide Falessi}, \bibinfo{person}{Clemente Izurieta}, \bibinfo{person}{Carolyn Seaman}, {and} \bibinfo{person}{Rodrigo Spínola}.} \bibinfo{year}{2023}\natexlab{b}.
\newblock \showarticletitle{Software practitioners’ point of view on technical debt payment}.
\newblock \bibinfo{journal}{\emph{Journal of Systems and Software}}  \bibinfo{volume}{196} (\bibinfo{year}{2023}), \bibinfo{pages}{111554}.
\newblock
\showISSN{0164-1212}
\href{https://doi.org/10.1016/j.jss.2022.111554}{doi:\nolinkurl{10.1016/j.jss.2022.111554}}


\bibitem[Gama et~al\mbox{.}(2020)]%
        {Gama2020}
\bibfield{author}{\bibinfo{person}{Eliakim Gama}, \bibinfo{person}{S{\'a}vio Freire}, \bibinfo{person}{Manoel Mendon{\c{c}}a}, \bibinfo{person}{Rodrigo~O Sp{\'\i}nola}, \bibinfo{person}{Matheus Paixao}, {and} \bibinfo{person}{Mariela~I Cort{\'e}s}.} \bibinfo{year}{2020}\natexlab{}.
\newblock \showarticletitle{Using Stack Overflow to Assess Technical Debt Identification on Software Projects}. In \bibinfo{booktitle}{\emph{34th Brazilian Symposium on Software Engineering (SBES)}}. \bibinfo{pages}{730--739}.
\newblock


\bibitem[Garousi et~al\mbox{.}(2020)]%
        {Garousi2020GreyLiterature}
\bibfield{author}{\bibinfo{person}{Vahid Garousi}, \bibinfo{person}{Michael Felderer}, \bibinfo{person}{Mika~V. Mäntylä}, {and} \bibinfo{person}{Aurona Rainer}.} \bibinfo{year}{2020}\natexlab{}.
\newblock \showarticletitle{Benefitting from the Grey Literature in Software Engineering Research}.
\newblock In \bibinfo{booktitle}{\emph{Contemporary Empirical Methods in Software Engineering}}. \bibinfo{publisher}{Springer International Publishing}, \bibinfo{address}{Cham}, \bibinfo{pages}{385--413}.
\newblock
\href{https://doi.org/10.1007/978-3-030-32489-6_14}{doi:\nolinkurl{10.1007/978-3-030-32489-6_14}}


\bibitem[Gomes et~al\mbox{.}(2023)]%
        {gomes2023investigating}
\bibfield{author}{\bibinfo{person}{Felipe Gomes}, \bibinfo{person}{Eder Santos}, \bibinfo{person}{S{\'a}vio Freire}, \bibinfo{person}{Thiago~Souto Mendes}, \bibinfo{person}{Manoel Mendon{\c{c}}a}, {and} \bibinfo{person}{Rodrigo Sp{\'\i}nola}.} \bibinfo{year}{2023}\natexlab{}.
\newblock \showarticletitle{Investigating the Point of View of Project Management Practitioners on Technical Debt-A Study on Stack Exchange}.
\newblock \bibinfo{journal}{\emph{Journal of Software Engineering Research and Development}} \bibinfo{volume}{11}, \bibinfo{number}{1} (\bibinfo{year}{2023}), \bibinfo{pages}{12--1}.
\newblock


\bibitem[Gomes et~al\mbox{.}(2022)]%
        {felipeUFBA}
\bibfield{author}{\bibinfo{person}{Felipe Gomes}, \bibinfo{person}{Eder Pereira~dos Santos}, \bibinfo{person}{S\'{a}vio Freire}, \bibinfo{person}{Manoel Mendon\c{c}a}, \bibinfo{person}{Thiago~Souto Mendes}, {and} \bibinfo{person}{Rodrigo Sp\'{\i}nola}.} \bibinfo{year}{2022}\natexlab{}.
\newblock \showarticletitle{Investigating the Point of View of Project Management Practitioners on Technical Debt: A Preliminary Study on Stack Exchange}. In \bibinfo{booktitle}{\emph{Proceedings of the International Conference on Technical Debt}} (Pittsburgh, Pennsylvania) \emph{(\bibinfo{series}{TechDebt '22})}. \bibinfo{publisher}{Association for Computing Machinery}, \bibinfo{address}{New York, NY, USA}, \bibinfo{pages}{31–40}.
\newblock
\showISBNx{9781450393041}
\href{https://doi.org/10.1145/3524843.3528095}{doi:\nolinkurl{10.1145/3524843.3528095}}


\bibitem[Gren and Ralph(2022)]%
        {gren2022makes}
\bibfield{author}{\bibinfo{person}{Lucas Gren} {and} \bibinfo{person}{Paul Ralph}.} \bibinfo{year}{2022}\natexlab{}.
\newblock \showarticletitle{What makes effective leadership in agile software development teams?}. In \bibinfo{booktitle}{\emph{Proceedings of the 44th international conference on software engineering}}. \bibinfo{pages}{2402--2414}.
\newblock


\bibitem[Gren et~al\mbox{.}(2017)]%
        {gren2017group}
\bibfield{author}{\bibinfo{person}{Lucas Gren}, \bibinfo{person}{Richard Torkar}, {and} \bibinfo{person}{Robert Feldt}.} \bibinfo{year}{2017}\natexlab{}.
\newblock \showarticletitle{Group development and group maturity when building agile teams: A qualitative and quantitative investigation at eight large companies}.
\newblock \bibinfo{journal}{\emph{Journal of Systems and Software}}  \bibinfo{volume}{124} (\bibinfo{year}{2017}), \bibinfo{pages}{104--119}.
\newblock


\bibitem[Hopkins(2016)]%
        {hopkins2016}
\bibfield{author}{\bibinfo{person}{Will~G. Hopkins}.} \bibinfo{year}{2016}\natexlab{}.
\newblock \bibinfo{booktitle}{\emph{A New View of Statistics}}.
\newblock
\urldef\tempurl%
\url{http://www.sportsci.org/resource/stats/}
\showURL{%
\tempurl}


\bibitem[House and Mitchell(1975)]%
        {house1975path}
\bibfield{author}{\bibinfo{person}{Robert~J House} {and} \bibinfo{person}{Terence~R Mitchell}.} \bibinfo{year}{1975}\natexlab{}.
\newblock \bibinfo{booktitle}{\emph{Path-goal theory of leadership}}.
\newblock \bibinfo{type}{{T}echnical {R}eport}.
\newblock


\bibitem[Kalliamvakou et~al\mbox{.}(2017)]%
        {kalliamvakou2017makes}
\bibfield{author}{\bibinfo{person}{Eirini Kalliamvakou}, \bibinfo{person}{Christian Bird}, \bibinfo{person}{Thomas Zimmermann}, \bibinfo{person}{Andrew Begel}, \bibinfo{person}{Robert DeLine}, {and} \bibinfo{person}{Daniel~M German}.} \bibinfo{year}{2017}\natexlab{}.
\newblock \showarticletitle{What makes a great manager of software engineers?}
\newblock \bibinfo{journal}{\emph{IEEE Transactions on Software Engineering}} \bibinfo{volume}{45}, \bibinfo{number}{1} (\bibinfo{year}{2017}), \bibinfo{pages}{87--106}.
\newblock


\bibitem[Khattak et~al\mbox{.}(2025)]%
        {khattak2025unwrapping}
\bibfield{author}{\bibinfo{person}{Shoukat~Iqbal Khattak}, \bibinfo{person}{Muhammad~Anwar Khan}, \bibinfo{person}{Muhammad~Iftikhar Ali}, {and} \bibinfo{person}{Abdul~Samad Kakar}.} \bibinfo{year}{2025}\natexlab{}.
\newblock \showarticletitle{Unwrapping IT Project Success in China: Examining the Role of Empowering Leadership, Workforce Agility and Top Management Support in IT Project Success}.
\newblock \bibinfo{journal}{\emph{SAGE Open}} \bibinfo{volume}{15}, \bibinfo{number}{1} (\bibinfo{year}{2025}), \bibinfo{pages}{21582440251318858}.
\newblock


\bibitem[Kitchenham et~al\mbox{.}(2023)]%
        {KitchenhamGrey}
\bibfield{author}{\bibinfo{person}{Barbara Kitchenham}, \bibinfo{person}{Lech Madeyski}, {and} \bibinfo{person}{David Budgen}.} \bibinfo{year}{2023}\natexlab{}.
\newblock \showarticletitle{How Should Software Engineering Secondary Studies Include Grey Material?}
\newblock \bibinfo{journal}{\emph{IEEE Transactions on Software Engineering}} \bibinfo{volume}{49}, \bibinfo{number}{2} (\bibinfo{year}{2023}), \bibinfo{pages}{872--882}.
\newblock
\href{https://doi.org/10.1109/TSE.2022.3165938}{doi:\nolinkurl{10.1109/TSE.2022.3165938}}


\bibitem[Li et~al\mbox{.}(2015)]%
        {li2015makes}
\bibfield{author}{\bibinfo{person}{Paul~Luo Li}, \bibinfo{person}{Amy~J Ko}, {and} \bibinfo{person}{Jiamin Zhu}.} \bibinfo{year}{2015}\natexlab{}.
\newblock \showarticletitle{What makes a great software engineer?}. In \bibinfo{booktitle}{\emph{2015 IEEE/ACM 37th IEEE International Conference on Software Engineering}}, Vol.~\bibinfo{volume}{1}. IEEE, \bibinfo{pages}{700--710}.
\newblock


\bibitem[Li et~al\mbox{.}(2012)]%
        {li2012leadership}
\bibfield{author}{\bibinfo{person}{Yan Li}, \bibinfo{person}{Chuan-Hoo Tan}, {and} \bibinfo{person}{Hock-Hai Teo}.} \bibinfo{year}{2012}\natexlab{}.
\newblock \showarticletitle{Leadership characteristics and developers’ motivation in open source software development}.
\newblock \bibinfo{journal}{\emph{Information \& Management}} \bibinfo{volume}{49}, \bibinfo{number}{5} (\bibinfo{year}{2012}), \bibinfo{pages}{257--267}.
\newblock


\bibitem[Mayring(2000)]%
        {Mayring_2000}
\bibfield{author}{\bibinfo{person}{Philipp Mayring}.} \bibinfo{year}{2000}\natexlab{}.
\newblock \showarticletitle{Qualitative Content Analysis}.
\newblock \bibinfo{journal}{\emph{Forum Qualitative Sozialforschung / Forum: Qualitative Social Research}} \bibinfo{volume}{1}, \bibinfo{number}{2} (\bibinfo{date}{Jun.} \bibinfo{year}{2000}).
\newblock
\href{https://doi.org/10.17169/fqs-1.2.1089}{doi:\nolinkurl{10.17169/fqs-1.2.1089}}


\bibitem[McHugh(2012)]%
        {McHugh2012}
\bibfield{author}{\bibinfo{person}{Mary~L McHugh}.} \bibinfo{year}{2012}\natexlab{}.
\newblock \showarticletitle{Interrater reliability: the kappa statistic}.
\newblock \bibinfo{journal}{\emph{Biochem Med (Zagreb)}} \bibinfo{volume}{22}, \bibinfo{number}{3} (\bibinfo{year}{2012}), \bibinfo{pages}{276--282}.
\newblock


\bibitem[Modi and Strode(2020)]%
        {modi2020leadership}
\bibfield{author}{\bibinfo{person}{Sunila Modi} {and} \bibinfo{person}{Diane Strode}.} \bibinfo{year}{2020}\natexlab{}.
\newblock \showarticletitle{Leadership in agile software development: a systematic literature review}.
\newblock  (\bibinfo{year}{2020}).
\newblock


\bibitem[Oliveira et~al\mbox{.}(2025)]%
        {oliveira2025investigating}
\bibfield{author}{\bibinfo{person}{Alessa Oliveira}, \bibinfo{person}{S{\'a}vio Freire}, \bibinfo{person}{Edna~Dias Canedo}, \bibinfo{person}{Manoel Mendon{\c{c}}a}, {and} \bibinfo{person}{Larissa Rocha}.} \bibinfo{year}{2025}\natexlab{}.
\newblock \showarticletitle{Investigating the Challenges Faced by Women on Software Engineering: a Grey Literature Study}. In \bibinfo{booktitle}{\emph{2025 IEEE/ACM Sixth Workshop on Gender Equality, Diversity, and Inclusion in Software Engineering (GEICSE)}}. IEEE, \bibinfo{pages}{17--24}.
\newblock


\bibitem[Papoutsoglou et~al\mbox{.}(2021)]%
        {papoutsoglou2021mining}
\bibfield{author}{\bibinfo{person}{Maria Papoutsoglou}, \bibinfo{person}{Johannes Wachs}, {and} \bibinfo{person}{Georgia~M Kapitsaki}.} \bibinfo{year}{2021}\natexlab{}.
\newblock \showarticletitle{Mining DEV for social and technical insights about software development}. In \bibinfo{booktitle}{\emph{2021 IEEE/ACM 18th International Conference on Mining Software Repositories (MSR)}}. IEEE, \bibinfo{pages}{415--419}.
\newblock


\bibitem[Piwowar-Sulej and Iqbal(2025)]%
        {piwowar2025sustainability}
\bibfield{author}{\bibinfo{person}{Katarzyna Piwowar-Sulej} {and} \bibinfo{person}{Qaisar Iqbal}.} \bibinfo{year}{2025}\natexlab{}.
\newblock \showarticletitle{Sustainability and software development projects: leadership, core self-evaluation and empowerment as critical success factors}.
\newblock \bibinfo{journal}{\emph{European Business Review}} \bibinfo{volume}{37}, \bibinfo{number}{2} (\bibinfo{year}{2025}), \bibinfo{pages}{371--394}.
\newblock


\bibitem[Procaccino et~al\mbox{.}(2005)]%
        {procaccino2005software}
\bibfield{author}{\bibinfo{person}{J~Drew Procaccino}, \bibinfo{person}{June~M Verner}, \bibinfo{person}{Katherine~M Shelfer}, {and} \bibinfo{person}{David Gefen}.} \bibinfo{year}{2005}\natexlab{}.
\newblock \showarticletitle{What do software practitioners really think about project success: an exploratory study}.
\newblock \bibinfo{journal}{\emph{Journal of Systems and Software}} \bibinfo{volume}{78}, \bibinfo{number}{2} (\bibinfo{year}{2005}), \bibinfo{pages}{194--203}.
\newblock


\bibitem[Ralph et~al\mbox{.}(2020)]%
        {ralph2020empirical}
\bibfield{author}{\bibinfo{person}{Paul Ralph}, \bibinfo{person}{Nauman bin Ali}, \bibinfo{person}{Sebastian Baltes}, \bibinfo{person}{Domenico Bianculli}, \bibinfo{person}{Jessica Diaz}, \bibinfo{person}{Yvonne Dittrich}, \bibinfo{person}{Neil Ernst}, \bibinfo{person}{Michael Felderer}, \bibinfo{person}{Robert Feldt}, \bibinfo{person}{Antonio Filieri}, \bibinfo{person}{Breno Bernard~Nicolau de França}, \bibinfo{person}{Carlo~Alberto Furia}, \bibinfo{person}{Greg Gay}, \bibinfo{person}{Nicolas Gold}, \bibinfo{person}{Daniel Graziotin}, \bibinfo{person}{Pinjia He}, \bibinfo{person}{Rashina Hoda}, \bibinfo{person}{Natalia Juristo}, \bibinfo{person}{Barbara Kitchenham}, \bibinfo{person}{Valentina Lenarduzzi}, \bibinfo{person}{Jorge Martínez}, \bibinfo{person}{Jorge Melegati}, \bibinfo{person}{Daniel Mendez}, \bibinfo{person}{Tim Menzies}, \bibinfo{person}{Jefferson Molleri}, \bibinfo{person}{Dietmar Pfahl}, \bibinfo{person}{Romain Robbes}, \bibinfo{person}{Daniel Russo}, \bibinfo{person}{Nyyti Saarimäki},
  \bibinfo{person}{Federica Sarro}, \bibinfo{person}{Davide Taibi}, \bibinfo{person}{Janet Siegmund}, \bibinfo{person}{Diomidis Spinellis}, \bibinfo{person}{Miroslaw Staron}, \bibinfo{person}{Klaas Stol}, \bibinfo{person}{Margaret-Anne Storey}, \bibinfo{person}{Davide Taibi}, \bibinfo{person}{Damian Tamburri}, \bibinfo{person}{Marco Torchiano}, \bibinfo{person}{Christoph Treude}, \bibinfo{person}{Burak Turhan}, \bibinfo{person}{Xiaofeng Wang}, {and} \bibinfo{person}{Sira Vegas}.} \bibinfo{year}{2020}\natexlab{}.
\newblock \bibinfo{title}{Empirical Standards for Software Engineering Research}.
\newblock
\showeprint[arxiv]{2010.03525}~[cs.SE]
\urldef\tempurl%
\url{https://arxiv.org/abs/2010.03525}
\showURL{%
\tempurl}


\bibitem[Rogers(2018)]%
        {rogers2018coding}
\bibfield{author}{\bibinfo{person}{Richard Rogers}.} \bibinfo{year}{2018}\natexlab{}.
\newblock \showarticletitle{Coding and writing analytic memos on qualitative data: A review of Johnny Salda{\~n}a's the coding manual for qualitative researchers}.
\newblock \bibinfo{journal}{\emph{The qualitative report}} \bibinfo{volume}{23}, \bibinfo{number}{4} (\bibinfo{year}{2018}), \bibinfo{pages}{889--893}.
\newblock


\bibitem[Salas et~al\mbox{.}(2005)]%
        {salas2005there}
\bibfield{author}{\bibinfo{person}{Eduardo Salas}, \bibinfo{person}{Dana~E Sims}, {and} \bibinfo{person}{C~Shawn Burke}.} \bibinfo{year}{2005}\natexlab{}.
\newblock \showarticletitle{Is there a “big five” in teamwork?}
\newblock \bibinfo{journal}{\emph{Small group research}} \bibinfo{volume}{36}, \bibinfo{number}{5} (\bibinfo{year}{2005}), \bibinfo{pages}{555--599}.
\newblock


\bibitem[Santana et~al\mbox{.}(2024)]%
        {santanaIEEESW}
\bibfield{author}{\bibinfo{person}{Beatriz Santana}, \bibinfo{person}{S\'{a}vio Freire}, \bibinfo{person}{Jos\'{e} Amancio~Macedo Santos}, {and} \bibinfo{person}{Manoel Mendon\c{c}a}.} \bibinfo{year}{2024}\natexlab{}.
\newblock \showarticletitle{Psychological Safety in the Software Work Environment}.
\newblock \bibinfo{journal}{\emph{IEEE Softw.}} \bibinfo{volume}{41}, \bibinfo{number}{4} (\bibinfo{date}{July} \bibinfo{year}{2024}), \bibinfo{pages}{86–94}.
\newblock
\showISSN{0740-7459}
\href{https://doi.org/10.1109/MS.2024.3386532}{doi:\nolinkurl{10.1109/MS.2024.3386532}}


\bibitem[Santana et~al\mbox{.}(2023)]%
        {santanaSBES}
\bibfield{author}{\bibinfo{person}{Beatriz Silva~De Santana}, \bibinfo{person}{S\'{a}vio Freire}, \bibinfo{person}{Leandro Cruz}, \bibinfo{person}{Lidiv\^{a}nio Monte}, \bibinfo{person}{Manoel Mendonca}, {and} \bibinfo{person}{Jos\'{e} Amancio~Macedo Santos}.} \bibinfo{year}{2023}\natexlab{}.
\newblock \showarticletitle{Exploring Psychological Safety in Software Engineering: Insights from Stack Exchange}. In \bibinfo{booktitle}{\emph{Proceedings of the XXXVII Brazilian Symposium on Software Engineering}} (Campo Grande, Brazil) \emph{(\bibinfo{series}{SBES '23})}. \bibinfo{publisher}{Association for Computing Machinery}, \bibinfo{address}{New York, NY, USA}, \bibinfo{pages}{503–513}.
\newblock
\showISBNx{9798400707872}
\href{https://doi.org/10.1145/3613372.3613411}{doi:\nolinkurl{10.1145/3613372.3613411}}


\bibitem[Sch{\"o}pfel and Farace(2010)]%
        {schopfel2010grey}
\bibfield{author}{\bibinfo{person}{Joachim Sch{\"o}pfel} {and} \bibinfo{person}{Dominic~J Farace}.} \bibinfo{year}{2010}\natexlab{}.
\newblock \showarticletitle{Grey literature}.
\newblock \bibinfo{journal}{\emph{Encyclopedia of library and information sciences}}  \bibinfo{volume}{3} (\bibinfo{year}{2010}), \bibinfo{pages}{2029--2039}.
\newblock


\bibitem[Spiegler et~al\mbox{.}(2021)]%
        {spiegler2021empirical}
\bibfield{author}{\bibinfo{person}{Simone~V Spiegler}, \bibinfo{person}{Christoph Heinecke}, {and} \bibinfo{person}{Stefan Wagner}.} \bibinfo{year}{2021}\natexlab{}.
\newblock \showarticletitle{An empirical study on changing leadership in agile teams}.
\newblock \bibinfo{journal}{\emph{Empirical Software Engineering}} \bibinfo{volume}{26}, \bibinfo{number}{3} (\bibinfo{year}{2021}), \bibinfo{pages}{41}.
\newblock


\bibitem[Strauss and Corbin(1998)]%
        {strauss1998basics}
\bibfield{author}{\bibinfo{person}{Anselm Strauss} {and} \bibinfo{person}{Juliet Corbin}.} \bibinfo{year}{1998}\natexlab{}.
\newblock \showarticletitle{Basics of qualitative research techniques}.
\newblock  (\bibinfo{year}{1998}).
\newblock


\bibitem[Subramanian and Banihashemi(2024)]%
        {subramanian2024towards}
\bibfield{author}{\bibinfo{person}{Salitha~Nair Subramanian} {and} \bibinfo{person}{Saeed Banihashemi}.} \bibinfo{year}{2024}\natexlab{}.
\newblock \showarticletitle{Towards modern leadership styles in the context of the engineering sector}.
\newblock \bibinfo{journal}{\emph{Project Leadership and Society}}  \bibinfo{volume}{5} (\bibinfo{year}{2024}), \bibinfo{pages}{100133}.
\newblock


\bibitem[Thite(1999)]%
        {thite1999leadership}
\bibfield{author}{\bibinfo{person}{Mohan Thite}.} \bibinfo{year}{1999}\natexlab{}.
\newblock \showarticletitle{Leadership: a critical success factor in IT project management}. In \bibinfo{booktitle}{\emph{PICMET'99: Portland International Conference on Management of Engineering and Technology. Proceedings Vol-1: Book of Summaries (IEEE Cat. No. 99CH36310)}}. IEEE, \bibinfo{pages}{298--303}.
\newblock


\bibitem[Wohlin et~al\mbox{.}(2012)]%
        {wohlin2012experimentation}
\bibfield{author}{\bibinfo{person}{Claes Wohlin}, \bibinfo{person}{Per Runeson}, \bibinfo{person}{Martin H{\"o}st}, \bibinfo{person}{Magnus~C Ohlsson}, \bibinfo{person}{Bj{\"o}rn Regnell}, \bibinfo{person}{Anders Wessl{\'e}n}, {et~al\mbox{.}}} \bibinfo{year}{2012}\natexlab{}.
\newblock \bibinfo{booktitle}{\emph{Experimentation in software engineering}}. Vol.~\bibinfo{volume}{236}.
\newblock \bibinfo{publisher}{Springer}.
\newblock


\bibitem[Xu et~al\mbox{.}(2024)]%
        {xu2024impact}
\bibfield{author}{\bibinfo{person}{Huimin Xu}, \bibinfo{person}{Meijun Liu}, \bibinfo{person}{Yi Bu}, \bibinfo{person}{Shujing Sun}, \bibinfo{person}{Yi Zhang}, \bibinfo{person}{Chenwei Zhang}, \bibinfo{person}{Daniel~E Acuna}, \bibinfo{person}{Steven Gray}, \bibinfo{person}{Eric Meyer}, {and} \bibinfo{person}{Ying Ding}.} \bibinfo{year}{2024}\natexlab{}.
\newblock \showarticletitle{The impact of heterogeneous shared leadership in scientific teams}.
\newblock \bibinfo{journal}{\emph{Information Processing \& Management}} \bibinfo{volume}{61}, \bibinfo{number}{1} (\bibinfo{year}{2024}), \bibinfo{pages}{103542}.
\newblock


\bibitem[Xu and Shen(2018)]%
        {xu2018role}
\bibfield{author}{\bibinfo{person}{Peng Xu} {and} \bibinfo{person}{Yide Shen}.} \bibinfo{year}{2018}\natexlab{}.
\newblock \showarticletitle{The role of leadership in agile software development}.
\newblock  (\bibinfo{year}{2018}).
\newblock


\end{thebibliography}
\end{document}